\documentclass[twocolumn,twocolappendix]{aastex63}

\received{}
\revised{}
\accepted{}
\submitjournal{ApJ}

\shorttitle{Shock in NLR powered by AGN outflows}
\shortauthors{Mizumoto et al.}
\graphicspath{{./}{figures/}}
\begin{document}

\title{Shock Excitation in Narrow Line Regions Powered by AGN Outflows}

\correspondingauthor{Misaki Mizumoto}
\email{mizumoto-m@fukuoka-edu.ac.jp}

\author[0000-0003-2161-0361]{Misaki Mizumoto}
\affil{Science education research unit, University of Teacher Education Fukuoka, Akama-bunkyo-machi, Munakata, Fukuoka 811-4192, Japan}

\author[0000-0001-6401-723X]{Hiroaki Sameshima}
\affiliation{Institute of Astronomy, Graduate School of Science, The University of Tokyo, Osawa, Mitaka, Tokyo 181-0015, Japan}

\author[0000-0003-4578-2619]{Naoto Kobayashi}
\affiliation{Kiso Observatory, Institute of Astronomy, School of Science, The University of Tokyo, Mitake, Kiso-machi, Kiso-gun, Nagano 397-0101, Japan}
\affiliation{Institute of Astronomy, Graduate School of Science, The University of Tokyo, Osawa, Mitaka, Tokyo 181-0015, Japan}
\affiliation{Laboratory of Infrared High-resolution Spectroscopy (LIH), Koyama Astronomical Observatory, Kyoto Sangyo University, Motoyama, Kamigamo, Kita-ku, Kyoto 603-8555, Japan}

\author{Noriyuki Matsunaga}
\affiliation{Department of Astronomy, Graduate School of Science, The University of Tokyo, Hongo, Bunkyo-ku, Tokyo, 113-0033, Japan}
\affiliation{Laboratory of Infrared High-resolution Spectroscopy (LIH), Koyama Astronomical Observatory, Kyoto Sangyo University, Motoyama, Kamigamo, Kita-ku, Kyoto 603-8555, Japan}

\author{Sohei Kondo}
\affiliation{Kiso Observatory, Institute of Astronomy, School of Science, The University of Tokyo, Mitake, Kiso-machi, Kiso-gun, Nagano 397-0101, Japan}

\author[0000-0002-6505-3395]{Satoshi Hamano}
\affiliation{National Astronomical Observatory of Japan, Osawa, Mitaka, Tokyo 181-8588, Japan}
\affiliation{Laboratory of Infrared High-resolution Spectroscopy (LIH), Koyama Astronomical Observatory, Kyoto Sangyo University, Motoyama, Kamigamo, Kita-ku, Kyoto 603-8555, Japan}

\author[0000-0003-3579-7454]{Chikako Yasui}
\affiliation{National Astronomical Observatory of Japan, Osawa, Mitaka, Tokyo 181-8588, Japan}

\author{Kei Fukue}
\affiliation{Laboratory of Infrared High-resolution Spectroscopy (LIH), Koyama Astronomical Observatory, Kyoto Sangyo University, Motoyama, Kamigamo, Kita-ku, Kyoto 603-8555, Japan}
\affiliation{Education Center for Medicine and Nursing, Shiga University of Medical Science, Seta Tsukinowa-cho, Otsu, Shiga 520-2192, Japan}

\author[0000-0002-5756-067X]{Akira Arai}
\affiliation{Subaru Telescope, National Astronomical Observatory of Japan, 650 North Aohoku Place, Hilo, Hawaii 96720, USA}

\author{Hideyo Kawakita}
\affiliation{Laboratory of Infrared High-resolution Spectroscopy (LIH), Koyama Astronomical Observatory, Kyoto Sangyo University, Motoyama, Kamigamo, Kita-ku, Kyoto 603-8555, Japan}
\affiliation{Department of Physics, Faculty of Science, Kyoto Sangyo University, Motoyama, Kamigamo, Kita-ku, Kyoto 603-8555, Japan}

\author{Shogo Otsubo}
\affiliation{Department of Physics, Faculty of Science, Kyoto Sangyo University, Motoyama, Kamigamo, Kita-ku, Kyoto 603-8555, Japan}
\affiliation{Laboratory of Infrared High-resolution Spectroscopy (LIH), Koyama Astronomical Observatory, Kyoto Sangyo University, Motoyama, Kamigamo, Kita-ku, Kyoto 603-8555, Japan}

\author[0000-0002-4896-8841]{Giuseppe Bono}
\affiliation{Dipartimento di Fisica, Universita di Roma Tor Vergata, via della Ricerca Scientifica 1, I-00133 Roma, Italy}
\affiliation{INAF-Osservatorio Astronomico di Roma, via Frascati 33, I-00078 Monte Porzio Catone, Italy}

\author[0000-0002-5878-5299]{Ivo Saviane}
\affiliation{European Southern Observatory, Alonso de Cordova 3107, Santiago, Chile}



\begin{abstract}
Outflows in the Active Galactic Nucleus (AGN)  are considered to  play a key role in the host galaxy evolution  through transfer of a large amount of energy. A Narrow Line Region (NLR) in the AGN is composed of ionized gas extending from pc-scales to kpc-scales.
It has been suggested that shocks are required for ionization of the NLR gas. If AGN outflows  generate such shocks, they will sweep  through the NLR  and the outflow energy will be transferred into  a galaxy-scale region. 
In order to study contribution of the AGN outflow to the NLR-scale shock,
we measure the [\ion{Fe}{2}]$\lambda12570$/[\ion{P}{2}]$\lambda11886$ line ratio, which is a good tracer of shocks, 
using near-infrared spectroscopic observations with WINERED (Warm INfrared Echelle spectrograph to Realize Extreme Dispersion and sensitivity) mounted on the New Technology Telescope. 
Among 13 Seyfert galaxies we observed, the [\ion{Fe}{2}] and [\ion{P}{2}] lines were detected in 12 and 6 targets, respectively. The [\ion{Fe}{2}]/[\ion{P}{2}] ratios in 4 targets were found to be higher than 10, which implies the existence of shocks.
We also found that the shock is likely to exist  where an ionized outflow, i.e., a blue wing in [\ion{S}{3}]$\lambda9533$, is present. 
Our result implies that the ionized outflow  present  over a  NLR-scale region sweeps  through the interstellar medium and  generates a shock.
\end{abstract}



\section{Introduction} \label{sec:intro}

The influence of the Active Galactic Nuclei (AGN) on the host galaxy evolution has been actively studied in recent decades (\citealt{har18} for review).
One example is the correlation between the mass of the central Super-Massive Black Hole (SMBH) and  velocity dispersion of the galactic spheroids (i.e., bulges and elliptical galaxies; \citealt{Msigma1, Msigma2}). 
The fact that there is a positive correlation between them implies their co-evolution, that is, SMBHs control the evolution of their host galaxies and vice versa. 
Theoretical studies  predict that 
AGN outflows, which originate from the radiation pressure due to accretion on to the black hole, offer ``a plausible physical origin'' \citep{kin15} for the co-evolution.
Impact of AGN on the host galaxy has also been implemented by many cosmological hydrodynamical simulations, such as the EAGLE project \citep{eagle3,eagle1,eagle2}, the SIMBA project \citep{simba19,simba19b,simba20}, and the IllustrisTNG project \citep{ITNG17,ITNG18,ITNG18b,ITNG19}.

The AGN outflows have a wide variety of velocities and size scales.
For example, in the innermost part ($\lesssim10^{-3}$~pc), UltraFast Outflows (UFO) with the velocity of 10--30\% of the light speed are observed in X-rays (e.g., \citealt{kin03b,pou03,ree03,tom10,tom14,gof13}). 
Their kinetic power usually exceeds the binding energy of the galactic spheroids by $\sim3$ magnitudes (see Equations 5 and 6 in \citealt{kin15}).
In the outer regions, multi-phase outflows are also observed; 
 one example is an ionized outflow, which is slower (several hundred km~s$^{-1}$) and more extended ($\sim$1~kpc) (e.g., \citealt{rup05,kom08,mul11,liu13,har14,kar16,Dav2020,sin22}).
 They are often observed as a blueshifted excess component (``blue wing''; \citealt{kom08}) in the optical [\ion{O}{3}]$\lambda5007$ line.
Recent studies have suggested that  they are driven by AGN radiation at $\sim1$~pc and travel up to $\sim1$~kpc \citep{wad18,mee21}.

In this work, we focus on physical processes in and around a Narrow Line Region (NLR).
The NLR extends from pc-scales to kpc-scales and emits forbidden lines with a line width of several hundred km s$^{-1}$ \citep{peterson}.
The main ionization mechanism in the NLR is photo-ionization from the AGN radiation \citep{ho93}.
However, it is known that  the photo-ionization cannot explain all the observed line features (e.g., \citealt{sim96}) and
 instead or in addition, shocks are  likely to  be responsible for some features (e.g., \citealt{kno96,wil99,mou00,rod04}).
Here, we came up with a scenario that a shock triggered by the AGN outflow contributes to the NLR ionization;
the fast velocity of the AGN outflow will generate the shock by interaction with the interstellar medium (ISM) (e.g., \citealt{richings2018,riching2018b}).
The shock may make a critical contribution to the NLR ionization, transferring energy from AGN to the NLR-scale cloud (i.e., the spheroid-scale cloud).
In these manners, the AGN outflow can carry its energy to the spheroid.
Some studies have partially supported this scenario; for example, \citet{Joh2021} found that
the gas density and velocity dispersion of the NLR region are  larger when the AGN is more active, which implies that NLR gas clouds are brought from the AGN outflows.

To investigate the existence of shocks in the NLR, 
\citet[T16]{ter16} studied the flux ratio of [\ion{Fe}{2}]$\lambda12570$ and [\ion{P}{2}]$\lambda11886$ emission lines observed in the near-infrared (NIR) $J$ band. 
\citet{oli01} proposed that the line flux ratio of [\ion{Fe}{2}] to [\ion{P}{2}] is a good tracer of shocks.
 The two lines share similar critical densities and ionization potentials, and are expected to be excited in similar physical environments. 
 By contrast, dust depletions are very different \citep{hob93}; iron is abundant in  dust whereas phosphorus is almost absent. When  a shock breaks  dust,  [\ion{Fe}{2}] appears much stronger than [\ion{P}{2}].
 Indeed, it is known that [\ion{Fe}{2}] is much  stronger in supernova remnants than [\ion{P}{2}], where shocks are present, whereas both have about the same flux in the Orion bar, where there are no known shocks \citep{oli01}.
T16 investigated the line ratios of 44 Seyfert galaxies (including some data taken from literature) and found that three of them (NGC 2782, NGC 5005, and Mrk 463) have  [\ion{Fe}{2}]/[\ion{P}{2}] ratios larger than 10,
 whereas more than half of them have a smaller one ($\sim2$).
They also found that the line ratios are not correlated with the radio loudness or starburst intensity, suggesting that the possible mechanism to trigger the shock is the AGN outflow.



The goal of this paper is to study the validity of our scenario that the AGN pc-scale outflow generate the shock in the NLR.
We describe our sample selection, observation setup, and data reduction in Section \ref{sec2}.
The obtained line profiles are explained in Section \ref{sec3}.
We discuss how the shocks  are  generated in NLRs in Section \ref{sec4} and summarize our  result in Section \ref{sec5}.

\section{Observations and data reduction}\label{sec2}
\subsection{Target selection and observations}

\begin{deluxetable*}{lcccccccc}
\tablecaption{Target list}\label{tab:targetlist}
\tablewidth{0pt}
\tablehead{
\colhead{Name} & \colhead{Redshift} & \colhead{Type}  & \colhead{$f_{\rm opt}$} & \colhead{$f_{\rm radio}$} & \colhead{$\log(R)$} & \colhead{$\log(\dot{M}_{\rm UFO})$} & \colhead{$\log(K_{\rm UFO})$} & \colhead{Reference}
\\
\colhead{} & \colhead{} &  \colhead{} &  \colhead{mJy} & \colhead{mJy} & \colhead{} &  \colhead{g s$^{-1}$}& \colhead{erg s$^{-1}$}& \colhead{} 
}
\decimalcolnumbers
\startdata
1H 0707--495& 0.0406 & Sy 1       &3 &$<1.6^*$ & $<-0.3$ &$\sim25.0$&$\sim44.3$&\citet{hag16}\\
Ark 120     & 0.0327  &Sy 1     &7.9 &3 &$-0.4$  &$>23.5$&$>43.1$&\citet{tom12}\\
ESO 323-G77 & 0.0150  &  Sy 2  & 16 & 36$^\dagger$ & 0.4 & $25.3\pm0.7$ & $42.1\pm0.7$ & \citet{tom12}\\
IC 4329A    &0.0161 & Sy 1&13 &67$^\dagger$ &0.7  &$>24.2$&$>42.8$&\citet{tom12}\\
IRAS 13224--3809 &0.0658 &  Sy 1&1.0 &5.4$^\dagger$ &0.8  &$24.8-25.5$&$44.4-44.8$&\citet{cha18}\\
MCG--5--23--16 & 0.0085 &Sy 2   &10 &6 &$-0.2$  &$23.9\pm1.0$&$42.7\pm1.0$&\citet{tom12} \\
MCG--6--30--15  & 0.0078  &Sy 1 &10 &1 & $-1.0$ &$23.8\pm0.3$&$40.1\pm0.3$&\citet{gof15}\\
NGC 1068   & 0.0038  &Sy 2      &0.6 &2.2 &0.6  &$24.1^{+0.2}_{-0.3}$&$43.6^{+0.2}_{-0.3}$&\citet{miz19}\\
NGC 1365    & 0.0055 & Sy 1    &350 &180 &$-0.3$  &$24.1\pm0.1$&$40.5\pm0.1$&\citet{gof15}\\
NGC 3783    & 0.0097&Sy 1   &40 &13 & $-0.5$ &$>24.2$&$<40.5$&\citet{gof15}\\
NGC 4507    & 0.0118 &Sy 2  &25 &11 & $-0.4$ &$>21.9$&$>41.2$&\citet{tom12}\\
NGC 6240    & 0.0243 &Sy 2   &20 &170 & 0.9 &$26.1^{+0.3}$&$44.9^{+0.3}$&\citet{miz19}\\
NGC 7582    & 0.0053 &Sy 2   &140 &110 &$-0.1$ &$23.8\pm1.0$&$43.4\pm1.1$&\citet{tom12}\\
\enddata
\tablecomments{
(1) Target name. (2) Redshift. (3) AGN type based on the optical spectra. Sy 1/2 means Seyfert 1/2 galaxy, respectively.
(4) Flux density at 4400~\AA\ \citep{Veron2010}. (5) Flux density at 5~GHz \citep{Veron2010} unless otherwise noted. $^*$Upper limit measured as a root mean square using the Giant Metrewave Radio Telescope (GMRT) 150~MHz all-sky radio survey data \citep{gmrt}.
$^\dagger$Flux density at 1.4~GHz. 
(6) Radio loudness ($f_{\rm radio}/f_{\rm opt}$).
(7) UFO mass loss rate.
(8) UFO kinetic power.
(9) References for (7) and (8).}
\end{deluxetable*}

\begin{deluxetable*}{llclclcc}
\tablecaption{Observation log}\label{tab:log}
\tablewidth{0pt}
\tablehead{
\colhead{Name} & \colhead{Date} & \colhead{Airmass} & \colhead{Mode}  &\colhead{Exposure} &\colhead{Std.\ star} &\colhead{Airmass of std.\ star} & \colhead{Seeing}\\
\colhead{} & \colhead{} & \colhead{} & \colhead{} &\colhead{sec$\times$freq.}  &\colhead{} &\colhead{} & \colhead{arcsec}
}
\decimalcolnumbers
\startdata
1H 0707$-$495    &2017 Nov 30 &1.07--1.08 &ABBA& $750\times4$ & 60 Ori & 1.15 & 1.48\\
Ark 120         &2017 Nov 30  &1.15--1.17 &ABBA & $900\times4$& 60 Ori & 1.15& 1.48 \\
ESO 323-G77    &2018 Mar 1   &1.03--1.05 &AB & $900\times2$& HD 139129& 1.09 & 0.93\\
IC 4329A        &2018 Mar 2   &1.05--1.07 &ABA& $300\times3$& HD 139129& 1.09&0.71 \\
IRAS 13224$-$3809&2018 Mar 2   &1.03--1.06 &ABBA ABBA & $900\times8$& HD 139129& 1.09 &0.71\\
MCG $-$5-23-16  &2018 Mar 1   &1.22--1.43 &ABBA & $900\times4$& HD 103101 & 1.42 & 0.84 \\
MCG $-$6-30-15  &2018 Mar 3    &1.01--1.02 &ABBA & $600\times4$&HD 118054 & 1.04 & 1.02 \\
NGC 1068        &2017 Nov 30  &1.16--1.20 &OSO& $900\times2$& 60 Ori & 1.15 & 1.48\\
NGC 1365        &2017 Nov 30  &1.02--1.10 &OSO OS & $900\times3$& 60 Ori & 1.15 & 1.48\\
NGC 3783        &2018 Feb 28  &1.15--1.20 &AB & $900\times2$& F Pup & 1.20 & 0.85\\
NGC 4507        &2018 Mar 3   &1.02--1.04 &OSO & $900\times2$& HD 118054 & 1.04 & 1.02 \\
NGC 6240        &2018 Mar 3   &1.31--1.38 &OSO &$900\times2$& 21 Oph & 1.23 & 0.77\\
NGC 7582        &2017 Nov 30 &1.07--1.11 &OSO & $900\times2$& 60 Ori & 1.15& 1.48\\
\enddata
\tablecomments{(1) Target name. (2) Observation date. (3) Airmass (minimum--maximum) (4) Observation mode. A/B shows the A-postion and B-position, and O/S shows the object and sky observation, respectively. (5) (Exposure time in sec for each snapshot)$\times$(the number of snapshots).
(6) Name of the telluric standard star. (7) Airmass of the tellluric standard star. (8) Seeing of the telluric standard star.
}
\end{deluxetable*}

\begin{deluxetable*}{lccccc}
\tablecaption{Pixel scale, spectral extraction region, and throw distance of each target}\label{tab:log2}
\tablewidth{0pt}
\tablehead{
\colhead{Name} & \colhead{Pixel scale} &
\multicolumn{3}{c}{Spectral extraction region} &
\colhead{Throw distance} \\
\colhead{}  & \colhead{pc/pixel}& \colhead{arcsec$\times$arcsec}& \colhead{pc$\times$pc}& \colhead{pix$\times$pix}&
\colhead{kpc}
}
\decimalcolnumbers
\startdata
1H 0707$-$495    &$2.2\times10^2$
&2.70$\times$1.08&$	(2.2\times10^3)	\times	(8.7\times10^2)	$&
$10\times4$ &9.8\\
Ark 120        &$1.8\times10^2$
&2.43$\times$1.08&$	(1.6\times10^3)	\times	(7.1\times10^2)	$&$9\times4$ &8.0 \\
ESO 323-G77    &$8.3\times10^1$
&2.43$\times$1.08&$	(7.5\times10^2)	\times	(3.3\times10^2)	$&$9\times4$ &3.7 \\
IC 4329A        &$8.9\times10^1$ 
&2.43$\times$1.08&$	(8.0\times10^2)	\times	(3.6\times10^2)	$&$9\times4$ &4.0\\
IRAS 13224$-$3809&$3.4\times10^2$
&2.70$\times$1.08&$	(3.4\times10^3)	\times	(1.4\times10^3)	$&$10\times4$ &15\\
MCG $-$5-23-16  &$4.7\times10^1$
&3.51$\times$1.08&$	(6.2\times10^2)	\times	(1.9\times10^2)	$&$13\times4$ &2.1  \\
MCG $-$6-30-15  &$4.4\times10^1$
&2.70$\times$1.08&$	(4.4\times10^2)	\times	(1.7\times10^2)	$&$10\times4$ &2.0 \\
NGC 1068      &$1.3\times10^1$ 
&2.97$\times$1.08&$	(1.4\times10^2)	\times	(5.2\times10^1)	$&$11\times4$ &---\\
NGC 1365       &$2.4\times10^1$
&3.24$\times$1.08&$	(2.8\times10^2)	\times	(9.4\times10^1)	$&$12\times4$ &---\\
NGC 3783       &$5.4\times10^1$
&2.97$\times$1.08&$	(6.0\times10^2)	\times	(2.2\times10^2)	$&$11\times4$ &2.4 \\
NGC 4507       &$6.6\times10^1$
&4.05$\times$1.08&$	(9.8\times10^2)	\times	(2.6\times10^2)	$&$15\times4$ &---  \\
NGC 6240       &$1.3\times10^2$
&2.70$\times$1.08&$	(1.3\times10^3)	\times	(5.3\times10^2)	$& $10\times4$ &---\\
NGC 7582       &$2.9\times10^1$
&4.32$\times$1.08&$	(4.6\times10^2)	\times	(1.2\times10^2)	$&$16\times4$ &---\\
\enddata
\tablecomments{(1) Target name. (2) Pixel scale. (3)--(5) Region used to extract the spectrum. (along the slit length)$\times$(slit width). (6) Distance between the centroids of A and B exposures. It corresponds to 45 pixels or 12.15 arcsec.
}
\end{deluxetable*}

The observations were carried out with the WINERED\footnote{The raw data on WINERED were generated at Kyoto Sangyo University. Derived data and relevant reduction scripts underlying this article are available from the corresponding author on reasonable request.} (Warm INfrared Echelle spectrograph to Realize Extreme Dispersion and sensitivity)  spectrograph \citep{winered,Ikeda2022} mounted on the New Technology Telescope (NTT) at the La Silla observatory as a visitor instrument.
WINERED has a bandwidth of 9100--13500~\AA. 
The slit length is 16.34 arcsec and the pixel scale is 0.27 arcsec/pixel.
The slit width was set to 1.08 arcsec (=4.0 pixel; WIDE mode) with $R=18000$ ($\Delta v=17$~km~s$^{-1}$).
The observations were carried out  for five nights between 2017 November 30 and 2018 March 3, with seeing of 0.7--1.5 arcsec.

The X-ray UFO catalog papers \citep{tom10,tom14,gof13} listed 36 objects. Within them, 11 objects can match our selection thresholds; they are located in the southern sky and have $J{\rm mag} <15.5$ and $z<0.07$, for which the spectral lines of interest ([\ion{Fe}{2}]$\lambda12570$ and  [\ion{P}{2}]$\lambda11886$) fall in the observable wavelength range.
In addition, two Seyfert galaxies, 1H 0707--495 and IRAS 13224$-$3809, have been reported to host strong UFOs \citep{hag16,par17}.
Thus we also observed them; the total number of targets is 13 \footnote{Although NGC 5506 also passed these criteria, we decided not to use it in this paper because of a defect in the observation data.}.
We also note that all of them are radio-quiet ($\log R<1$, where $R$ is the ratio of the flux density at 5~GHz to 4400~\AA), that is, we can exclude potential energy inputs from  radio jets and to study the influence of multi-scale outflows.
Tables \ref{tab:targetlist} and \ref{tab:log} show the target list and the observation log, respectively.

When the host galaxy emission  was anticipated to potentially occupy most of the slit length of 16.34~arcsec, 
we used an Object-Sky-Object mode (``OSO'' mode), that is, the blank sky was observed between the object observations.
Any bright sources were not observed  immediately before the  targeted observations to avoid persistence (see section 5.3 in \citealt{Ikeda2022}). 
Where necessary, we waited until the persistence gets sufficiently low before we start observing the targets.

When a target did not occupy the slit length, we adopt the ``ABBA'' mode.
In this mode we position the object at a location offset from the center of the slit (designated as A-position), and then, for the next exposure, position it at a location shifted in the opposite direction (designated as B-position). 
By subtracting the spectra obtained from these two positions (that is, A--B or B--A), the contribution of the sky background is removed and we get an image with two spectra, one positive and one negative.
In order to discuss NLR in AGN with kpc-scale spatial extent, only those with the throw distance (distance between the centroids of A and B exposures) more than 2~kpc are treated in this mode. 

Data reduction was performed using IRAF\footnote{IRAF is distributed by the National Optical Astronomy Observatories, which is operated by the Association of Universities for Research in Astronomy, Inc.\ (AURA) under cooperative agreement with the National Science Foundation.} and PyRAF\footnote{PyRAF is a product of the Space Telescope Science Institute, which is operated by AURA for NASA.}.

\begin{figure}
    \centering
    \includegraphics[width=5.5cm,angle=270]{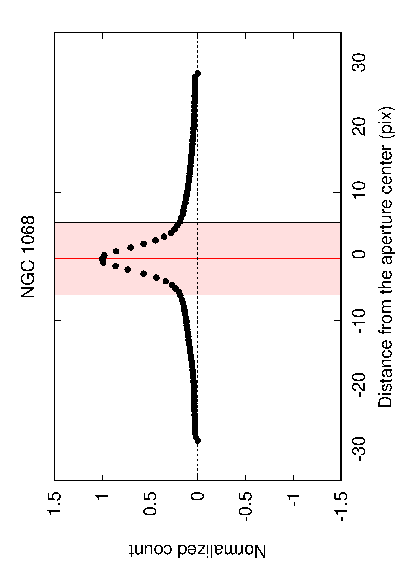}
    \includegraphics[width=5.5cm,angle=270]{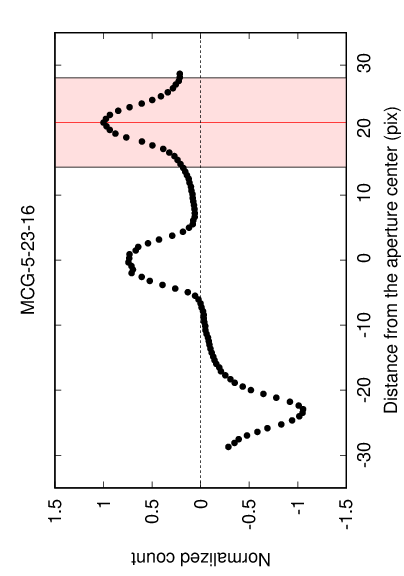}
    \caption{Spatial profiles of the emission along the slit.
    The upper panel shows the NGC 1068 case, in which the blank sky was observed between the object observations for the sky subtraction. The profile shows the one in the object observation. 
    The source region at the center is automatically selected with Gaussian fitting.
    The lower panel shows the MCG--5-23-16 case, in which the image of A-position was subtracted from the image of B-position and the one-dimensional spatial profile around the slit was calculated. The positive peak at a positive distance (right-hand side in the panel) is  the signal  at the A-position, whereas the negative one in the left hand is for B. The spectral extraction region (red) is automatically selected,  as in the upper panel. 
    The  peak at the center is the  remaining counts  from a bright telluric standard star observed shortly before this target observation.}
    \label{fig:2dmap}
\end{figure}

\begin{figure}
    \centering
    \includegraphics[width=7cm]{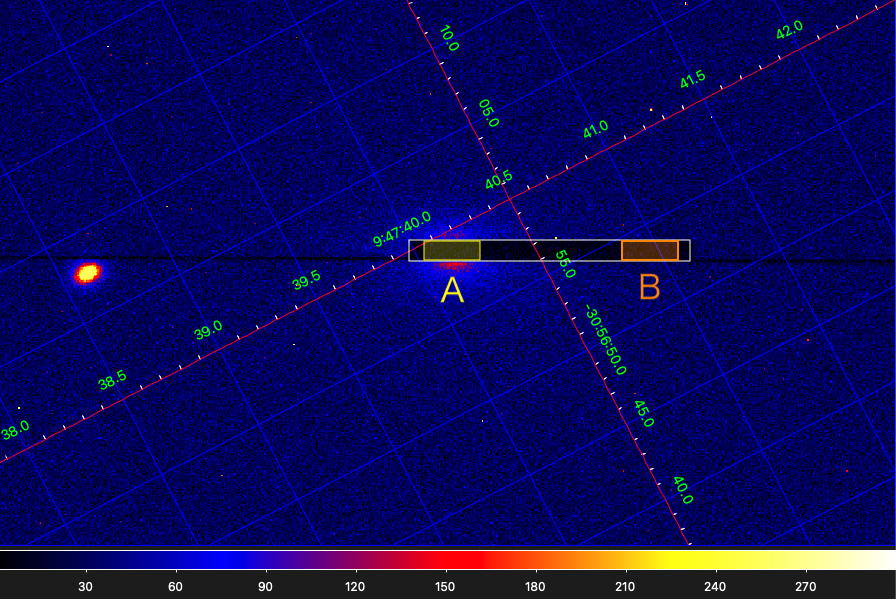}
    \caption{The slit viewer image of MCG--5-23-26 (A-position). Please note that the left and right sides are reversed from Figure \ref{fig:2dmap}. The white, yellow-shaded, and orange-shaded rectangles show the WINERED slit, the spectral extraction region (A-position), and the sky region (B-position), respectively. We note that the yellow-shaded rectangle corresponds to the red-shaded region in Figure \ref{fig:2dmap}. The signal on the left is a serendipitous point source (2MASS J09473861--3057050).}
    \label{fig:MCG5_sv}
\end{figure}

\subsection{Data reduction}
The detailed explanation of the data reduction (e.g., wavelength calibration, dispersion calibration, flat fielding) in WINERED are described in Section 8 in \citet{Ikeda2022}.
Hereafter data reduction unique to this paper are explained.

The source region was selected by the following steps.
First, a spatial profile along the slit length was plotted and the peak was searched (red vertical lines in Figure \ref{fig:2dmap}).
Next, the spatial profile around the peak was fitted with a single Gaussian to calculate a Full Width at Half maximum (FWHM).
Last, the source region was selected from $-1\times$FWHM to $+1\times$FWHM (red-shaded regions in Figures \ref{fig:2dmap}).
The sky spectrum was set to have the same width as the source one.
We used Skycorr \citep{skycorr} to subtract the sky spectrum under the situation that the infrared background can change substantially.
The sky-subtracted spectra were normalized to unity for the continuum, using a {\tt continuum} command in PyRAF.
In the {\tt continuum} command, the spectra were normalized by a cubic Legendre polynomial, without emission/absorption-line-dominant bands.
As a next step, telluric absorption lines were corrected by a {\tt telluric} command, using A0-type standard stars with similar airmass observed before or after a suit of observations (Table \ref{tab:log}). 
In the {\tt telluric} command, we determined the optimal value of the scale and wavelength shift of the telluric standard spectrum after removing intrinsic lines (see \citealt{sam18} for detailed method of removing the intrinsic lines) and divided it from the target spectrum.

It is imperative to prevent the extended signal of a galaxy from contaminating the sky background, which is then subtracted from each of the pair of A and B positions.
The targets presented in this study possess low redshift; therefore, 
it is imperative to prevent the extended signal of a galaxy from contaminating the sky background, which is then subtracted from each of the pair of A and B positions.
Column (5) in Table \ref{tab:log2} provides insights into the size of the spectral extraction region, a product of Gaussian fitting as illustrated in Figure \ref{fig:2dmap}. The spatial extension of the emission region along the slit length falls within the range of 9--13 pixels for the ``ABBA'' mode. Although it is worth noting that this spatial extent is determined using the NIR continuum and may not be entirely congruent with the NLR extent, which should ideally be assessed via narrow lines, we maintain confidence that over-subtraction was effectively mitigated in the A--B sky subtraction process. This confidence is supported by the fact that most of the emission resides within this defined region, and the spatial extent (9-13 pixels for A--B subtraction) remains significantly smaller than the throw distance of 45 pixels by a factor exceeding three. Figure \ref{fig:MCG5_sv} visually exemplifies the spectral extraction region and the sky region, reaffirming that extended emission in the A-position does not contaminate the B-position, and vice versa.

Since we are interested in the line intensity ratio and the shape of the line profile, we did not observe the flux standard star for measuring the absolute value of flux.
Then, all the exposure were stacked and the continuum of the total spectrum was normalized again.
As the last step, the barycentric correction was performed.

\section{Results}\label{sec3}
\subsection{The [\ion{Fe}{2}] and [\ion{P}{2}] lines}\label{sec3.1}

\begin{figure*}
\gridline{\rotatefig{0}{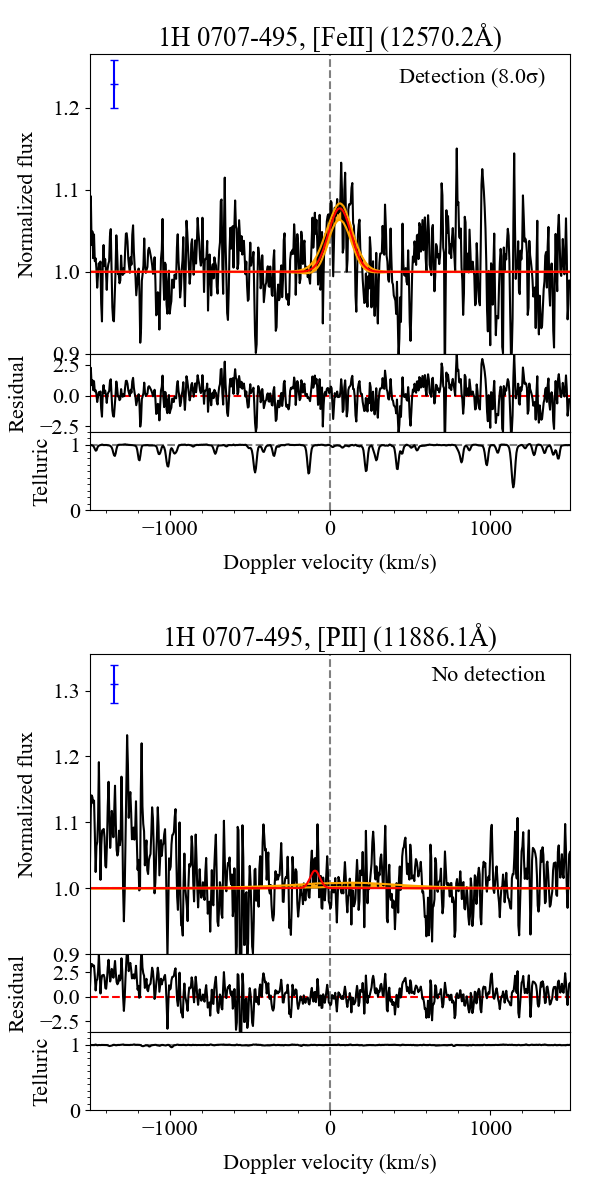}{0.25\textwidth}{(a)}
          \rotatefig{0}{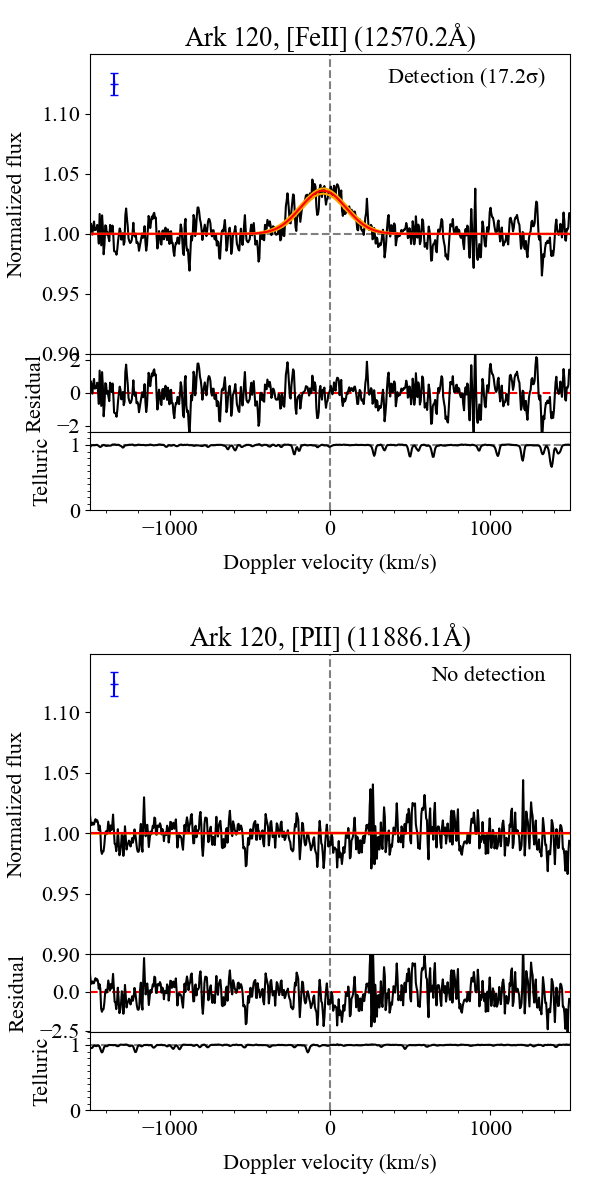}{0.25\textwidth}{(b)}
          \rotatefig{0}{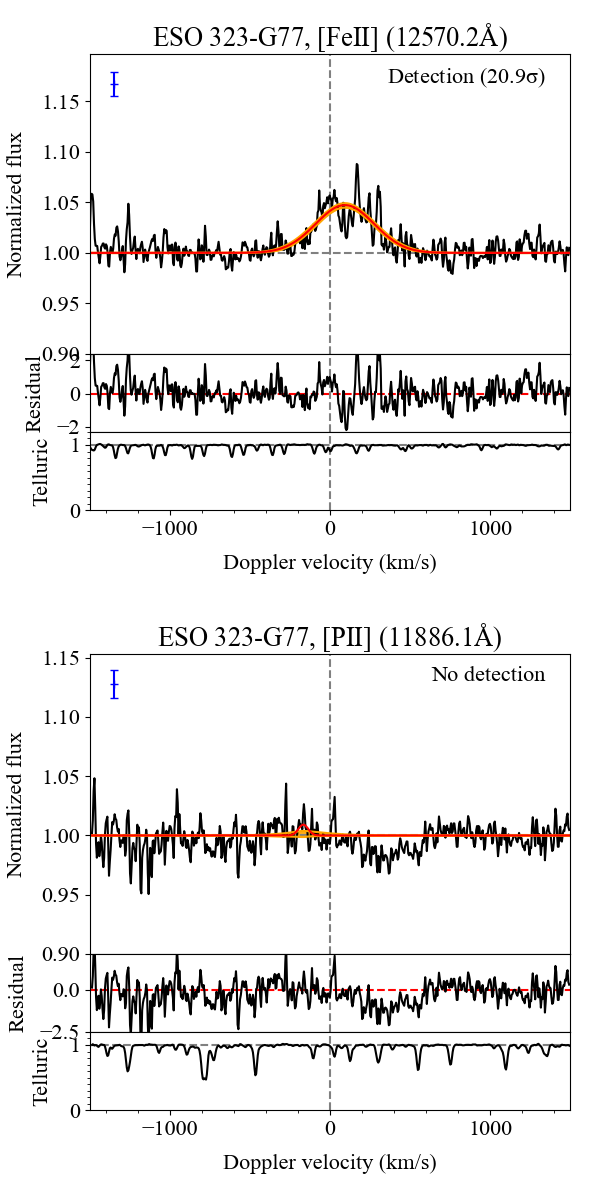}{0.25\textwidth}{(c)}
           \rotatefig{0}{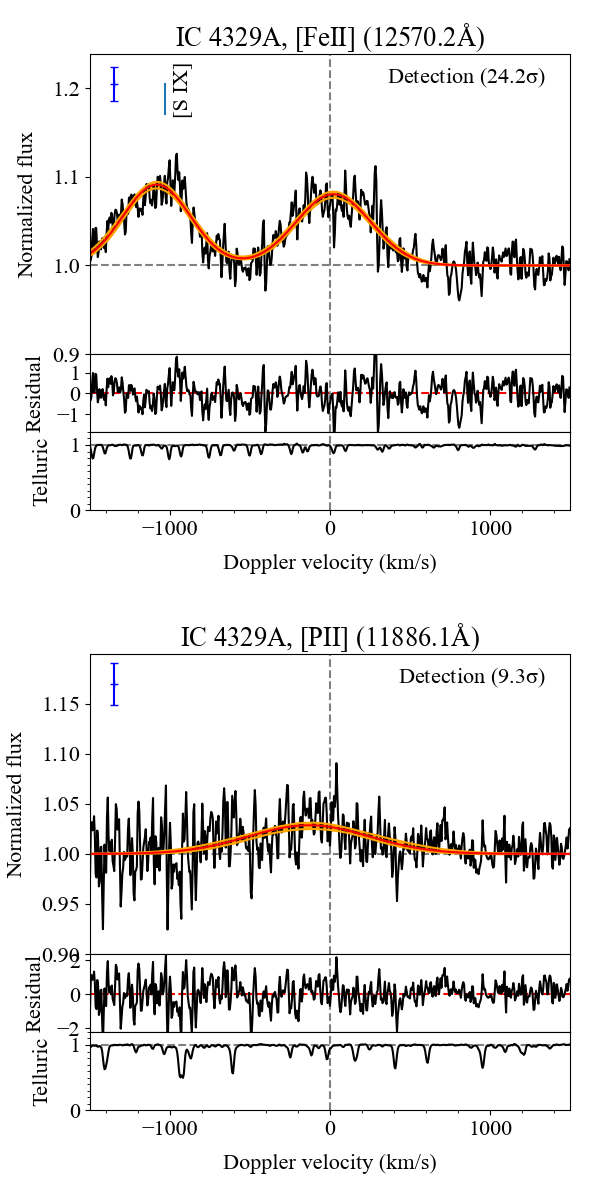}{0.25\textwidth}{(d)}
          }
\gridline{
           \rotatefig{0}{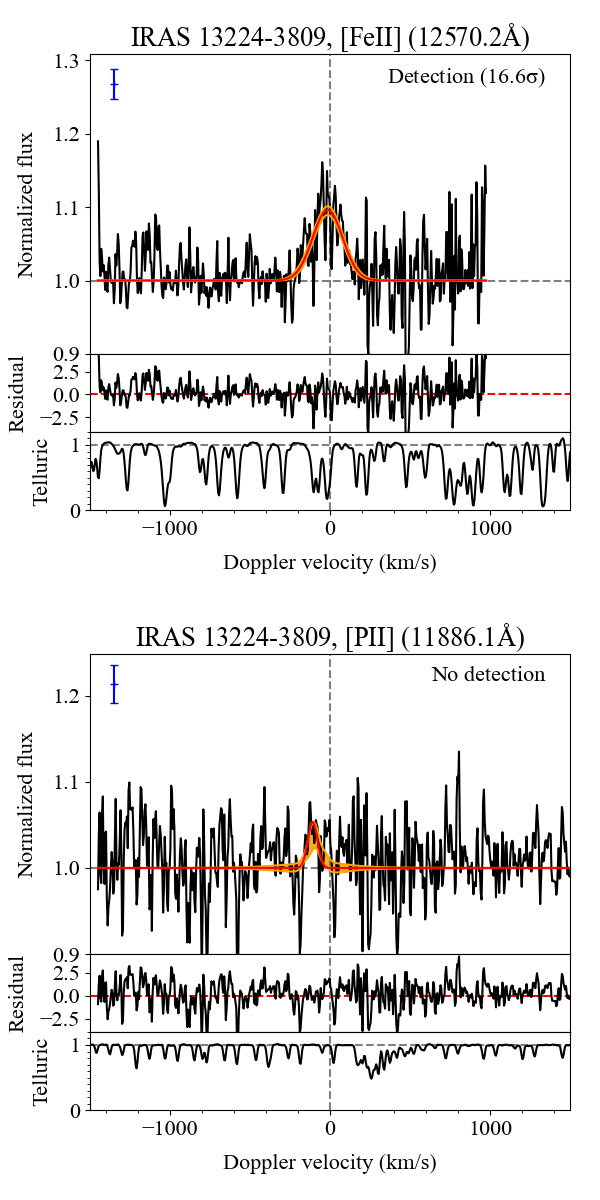}{0.25\textwidth}{(e)}
           \rotatefig{0}{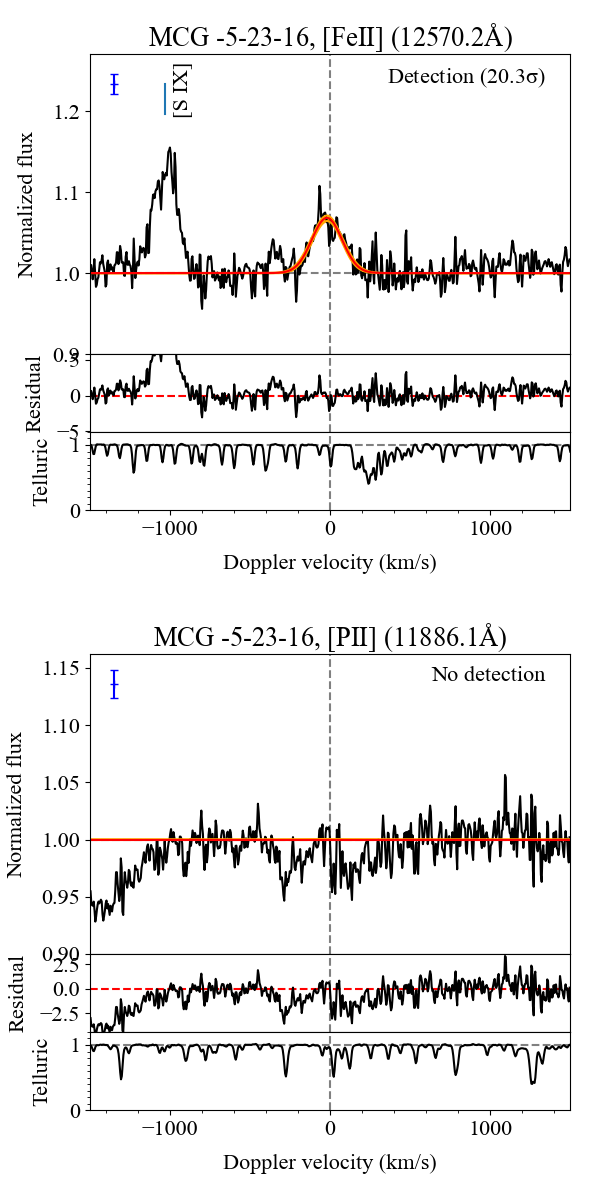}{0.25\textwidth}{(f)}
           \rotatefig{0}{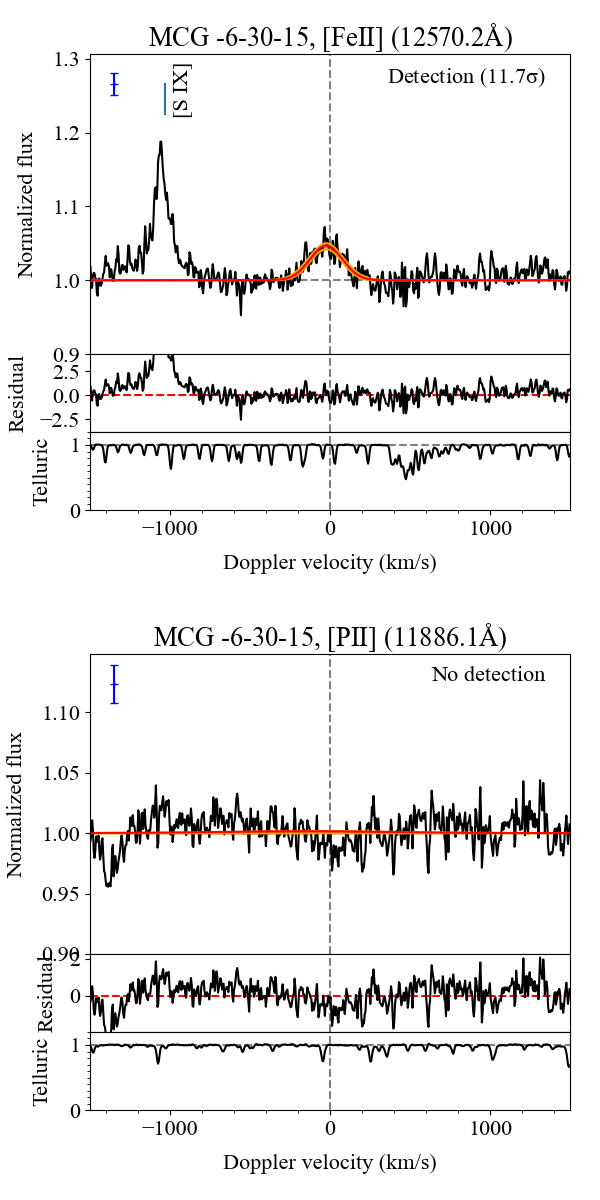}{0.25\textwidth}{(g)}
           \rotatefig{0}{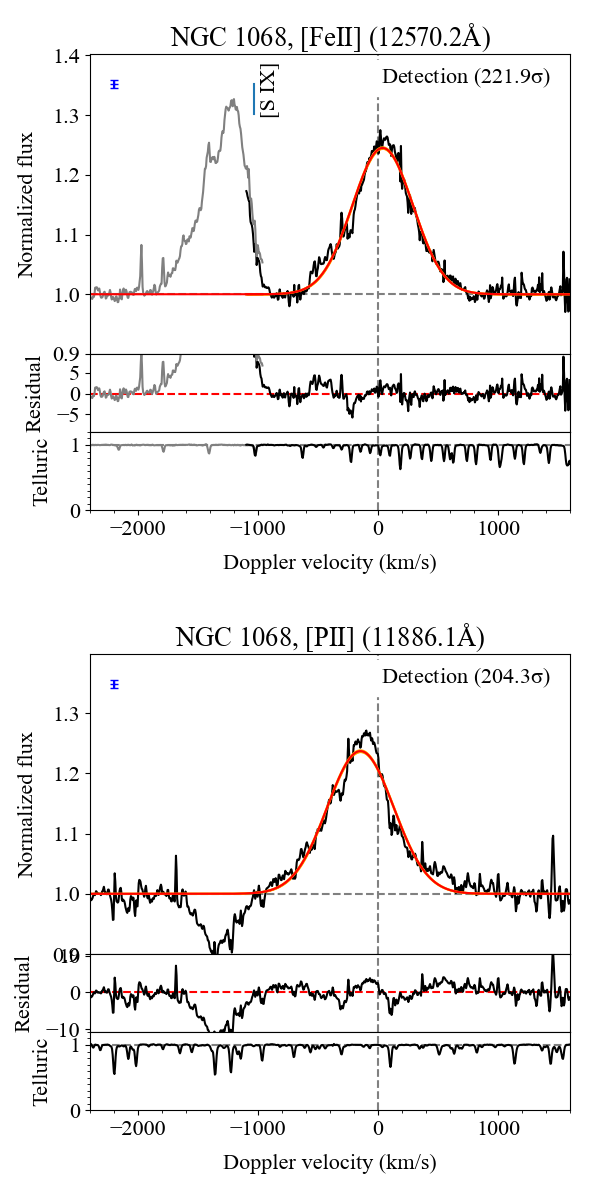}{0.25\textwidth}{(h)}
           } 
\caption{Velocity profiles of the [\ion{Fe}{2}] and [\ion{P}{2}] emission lines. Positive velocities mean redshift, and vice versa.  
The red line in each panel shows the model (continuum + Gaussian) when the parameter with the maximum likelihood in the sampling is adopted.
The orange shade shows the $1\sigma$ posterior spread from the sampling median for each wavelength bin.
The middle and bottom panels show the residuals and the telluric absorption, respectively. The blue bar at the upper left shows the typical error on the continuum. The gray shows the spectrum in the adjacent Echelle order. The adjacent [\ion{S}{9}] line is labeled. In the line detection case, its significance is written at the upper right.
(a) 1H 0707$-$495, (b) Ark 120, (c) ESO 323-G77, (d) IC 4329A, (e) IRAS 13224$-$3809, (f) MCG $-$5-23-16, (g) MCG $-$6-30-15, and (h) NGC 1068.}
\label{fig:lineprofile}
\end{figure*}

\addtocounter{figure}{-1}
\begin{figure*}
\gridline{
           \rotatefig{0}{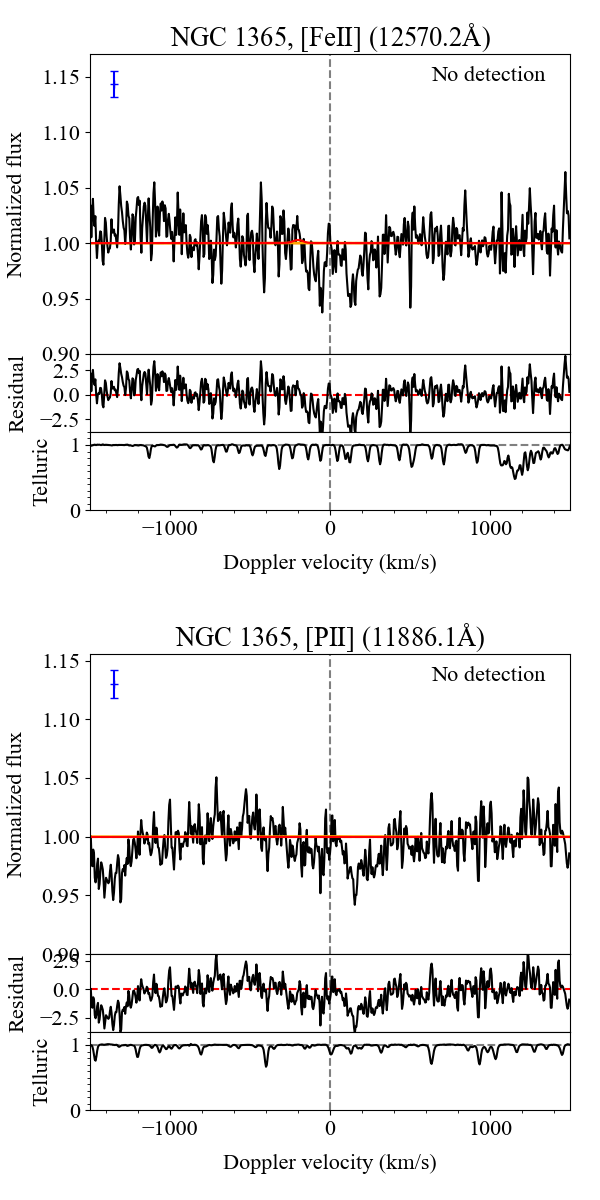}{0.25\textwidth}{(i)}
           \rotatefig{0}{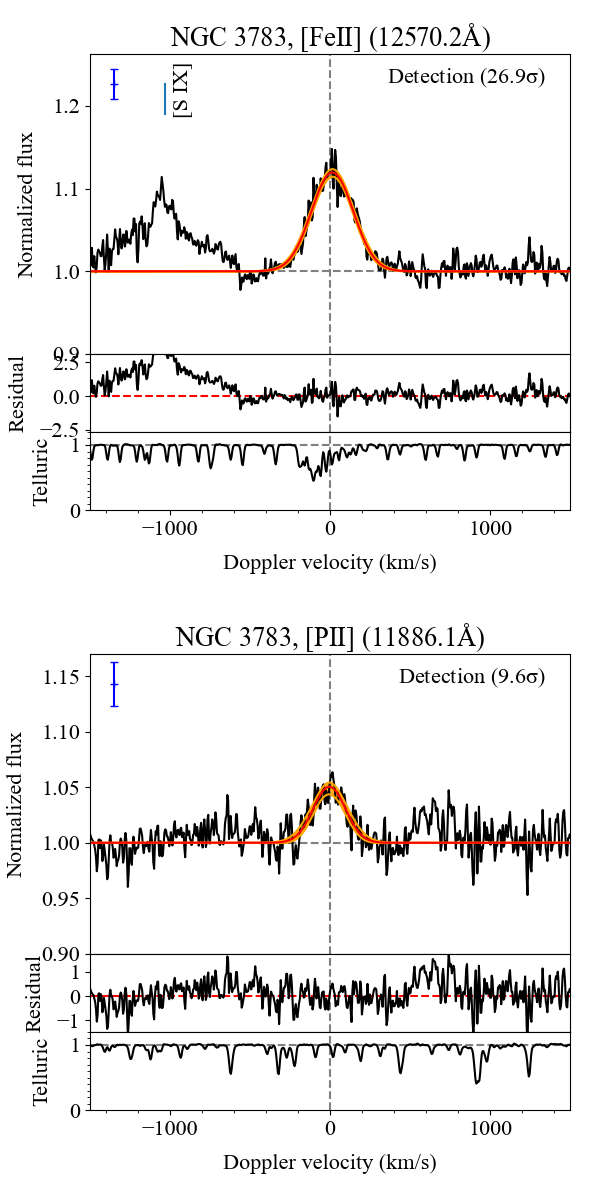}{0.25\textwidth}{(j)}
           \rotatefig{0}{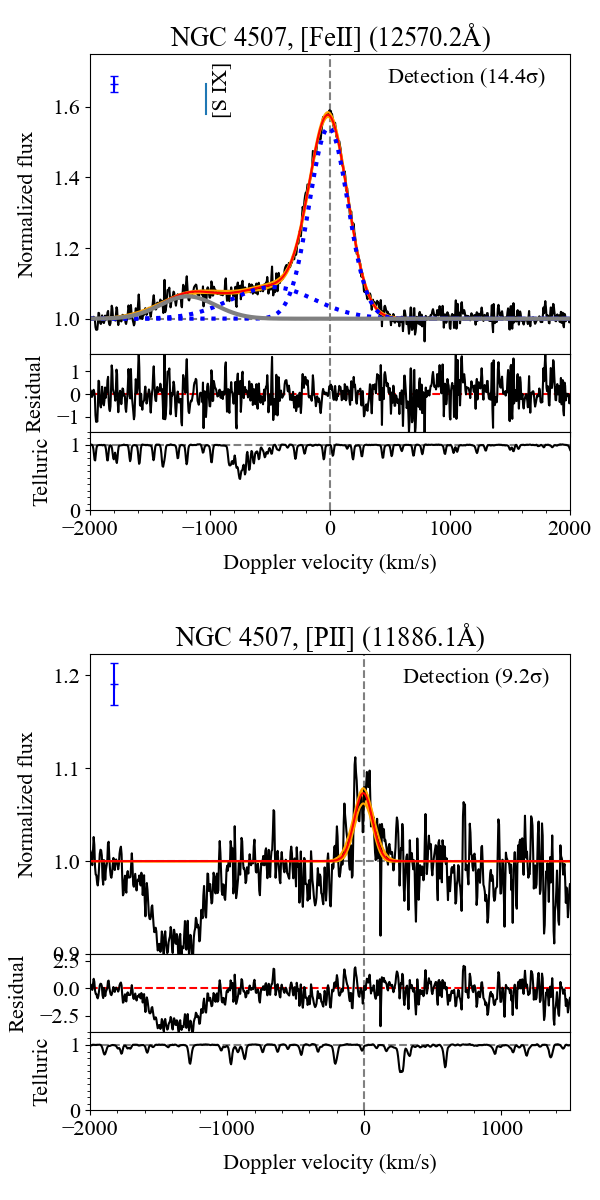}{0.25\textwidth}{(k)}
           \rotatefig{0}{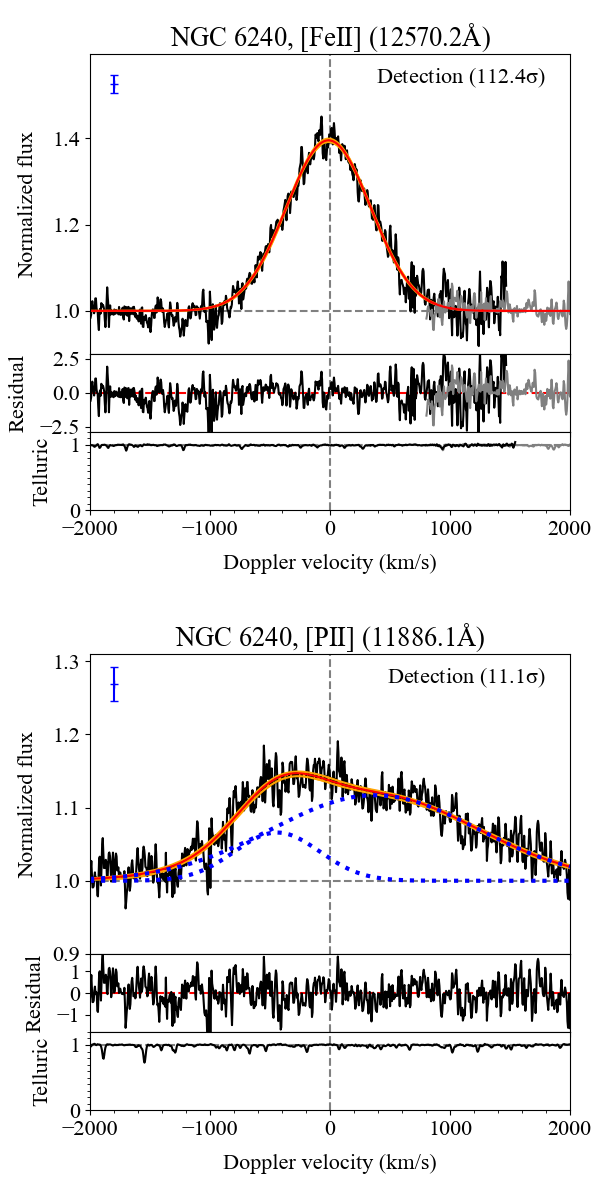}{0.25\textwidth}{(l)}
          }
\gridline{\leftfig{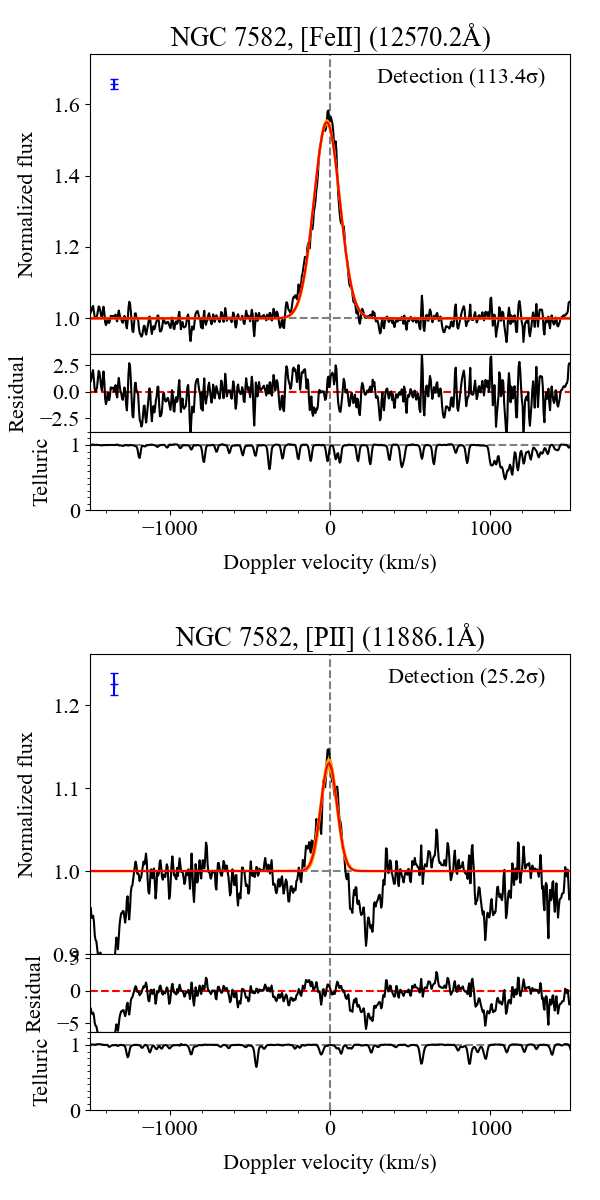}{0.25\textwidth}{(m)}
          }
          \caption{{\it Continued.}
          In NGC 4507, the adjacent [\ion{S}{9}] line is introduced as the gray solid line.
(i) NGC 1365, (j) NGC 3783, (k) NGC 4507, (l) NGC 6240, and (m) NGC 7582. When an additional Gaussian is necessary to introduce, each Gaussian component is shown in the blue dotted line.}
\label{fig:lineprofile2}
\end{figure*}

The obtained line profiles of [\ion{Fe}{2}] and [\ion{P}{2}]  are shown in Figure \ref{fig:lineprofile}.
The line fitting was performed by the Markov chain Monte Carlo methods (MCMC) using the emcee module \citep{emcee} in Python.
We set the number of independent chains (walkers) exploring the parameter space (nwalker) to be 512, and 
the number of iterations performed by each walker in the MCMC process (niter) to be 3000. 
The fitting procedure is as follows:
(1) One positive Gaussian was introduced. The line width (standard deviation; $\sigma_v$) was set to be larger than the velocity resolution of 17~km~s$^{-1}$. Because the line center may be slightly shifted due to gas motion, the central velocity was set to be free within $\pm200$~km/s from the rest frame.
We summarize the line intensity in Table \ref{tab:result} as an equivalent width (EW).
Its error ($1\sigma$) was calculated from the Bayesian posterior probability (See Figure \ref{fig:contour} in Appendix \ref{sec:contour} for corner plots of the MCMC fitting).
The adjacent [\ion{S}{9}] emission line was simultaneously modeled for the fitting of [\ion{Fe}{2}] in IC 4329A, because the fitting model was distorted without it.
(2) If the line intensity did not exceed three times its error, the line was considered undetectable and the three times of the error was adopted as the upper limit.
(3) If some residuals were seen after introducing the first Gaussian and they were overlapped with the line of interest, additional Gaussians were introduced ([\ion{Fe}{2}] in NGC 4507 and [\ion{P}{2}] in NGC 6240). To determine whether the additional Gaussian was necessary, an F test \citep{ftest1,ftest2} was performed based on the chi squared and degrees of freedom before and after the Gaussian was added. The additional one was introduced only when it was judged to be necessary at a 99.7\% confidence level. This procedure was not performed when the residuals are apart enough to be unaffected by fitting of the line of interest (e.g., [\ion{Fe}{2}] in NGC 3783).

The EW ratios of [\ion{Fe}{2}] to [\ion{P}{2}], as well as the obtained EW and $\sigma_v$ of each line, are listed in Table \ref{tab:result}.
We detected the [\ion{Fe}{2}] line  in 12 out of 13 AGNs and [\ion{P}{2}] in 6 of them.
Neither of the lines was detected in NGC 1365.

While T16 calculated the [\ion{Fe}{2}]/[\ion{P}{2}] ratios using the line fluxes,
we use the EWs.
Our ratios must be $F_{\lambda,\,{\rm [Fe\,II]}}/F_{\lambda,\,{\rm [P\,II]}}$ times larger than those of T16, where $F_{\lambda,\,k}$ refers to the continuum flux density at the $k$ line.
The NIR power-law slope in the wavelength unit ($\alpha_\lambda$, where $F_\lambda\propto\lambda^{\alpha_\lambda}$) in the quasar composite spectrum is known to be $\alpha_\lambda=-0.19$ \citep{gli06}.
If the slope in Seyfert galaxies is similar to those in quasars, the factor is  $F_{\rm \lambda,\,[Fe\,II]}/F_{\rm \lambda,\,[P\,II]}=(1.257/1.188)^{-0.19}=0.99$.
Therefore the difference of the line ratio definitions by us and T16 hardly affects the results.

We found that four sources (Ark 120, ESO 323-G77, MCG $-$5-23-16, and NGC 4507) have the [\ion{Fe}{2}]/[\ion{P}{2}] ratios  larger than 10; notably, 
for NGC 4507, [\ion{P}{2}] was detected and the ratio exceeded 10, which is the first report.
They are likely to have shocks in the NLR.
This result doubles the number of the known AGN samples with the line ratio  larger than 10, where the most major previous work was done by T16.

\begin{splitdeluxetable*}{lcccccBccccc}
\tablecaption{Results of the NTT/WINERED observations \label{tab:result}}
\tablewidth{0pt}
\tablehead{
\colhead{Name} & \colhead{[\ion{Fe}{2}] EW} & \colhead{[\ion{Fe}{2}] $\sigma_v$} & \colhead{[\ion{P}{2}] EW} & \colhead{[\ion{P}{2}] $\sigma_v$} & \colhead{[\ion{Fe}{2}]/[\ion{P}{2}]}
& \colhead{[\ion{S}{3}]$_{\rm peak}$ EW} 
  & \colhead{[\ion{S}{3}]$_{\rm peak}$ $\sigma_v$}& \colhead{[\ion{S}{3}]$_{\rm wing}$ EW} & \colhead{[\ion{S}{3}]$_{\rm wing}$ velocity} 
  & \colhead{[\ion{S}{3}]$_{\rm wing}$ $\sigma_v$}\\
\colhead{}  &  \colhead{\AA}&  \colhead{km s$^{-1}$} &\colhead{\AA} &  \colhead{km s$^{-1}$}  & \colhead{}  & \colhead{\AA} & \colhead{km s$^{-1}$} & \colhead{\AA} & \colhead{km s$^{-1}$}& \colhead{km s$^{-1}$} 
}
\decimalcolnumbers
\startdata
1H 0707$-$495 
& $0.58\pm0.07$ & $75^{+13}_{-10}$ & $<0.32$ & --- & $>1.8$ 
& $2.6\pm0.2$ & --- & ---&---&---\\
Ark 120& $0.54\pm0.03$ & $144\pm9$ & $<0.027$ & --- & $>20.0$ 
& $3.8\pm0.5$ & $173\pm6$ & $0.85\pm0.52$ & $-310\pm90$ & $200\pm50$ \\
ESO 323-G77& $0.91\pm0.04$ & $184^{+10}_{-9}$ & $<0.075$ & --- & $>12.2$ 
& $6.0\pm0.2$ & $200\pm10$ &--- & --- & --- \\
IC4329A 
& $2.04\pm0.14$ & $240\pm10$ & $1.09\pm0.12$ & $390\pm50$ & $1.87\pm0.24$ &
 $26.4\pm0.2$ & $235\pm2$ & --- & --- &  --- \\
IRAS 13224$-$3809& $0.95\pm0.06$ & $94\pm6$ & $<0.20$ & --- & $>4.9$ 
& $1.17\pm0.06$ & $55\pm4$ & $7.1\pm1.2$ & $-217\pm11$ & $417\pm9$ \\
MCG $-$5-23-16 & $0.68\pm0.03$ & $95^{+6}_{-5}$ & $<0.012$ & --- & $>56.5$ 
& $13.47\pm0.06$ & $83.5\pm0.4$ & $0.31\pm0.02$ & $-292\pm3$ & $43\pm3$ \\
MCG $-$6-30-15 
& $0.50\pm0.04$ & $100\pm10$ & $<0.12$ & --- & $>4.1$ 
& $9.81\pm0.06$ & $60\pm2$ & --- & --- &  --- \\
NGC 1068
& $6.53\pm0.03$ & $253\pm1$ & $6.53\pm0.03$ & $277\pm2$ & $1.00\pm0.01$ 
& $170.2\pm0.3$ & --- & $6.91\pm0.04$ & $-860\pm10$ & $144.7\pm0.7$\\ 
NGC 1365& $<0.018$ & --- & $<0.015$ & --- & --- 
& $10.4\pm0.4$ & --- & --- & --- &  --- \\
NGC 3783& $1.65\pm0.06$ & $131^{+6}_{-5}$ & $0.51\pm0.05$ & $104^{+13}_{-11}$ & $3.23\pm0.36$ 
& $19.2\pm0.8$ & --- & --- & --- &  --- \\
NGC 4507
& $12.1\pm0.8$ & --- & $0.49\pm0.05$ & $70^{+10}_{-9}$ & $25\pm3$ 
& $35.5\pm0.7$ & $137\pm3$ & $22.3\pm0.3$ & $-100\pm20$ & $400\pm100$ \\
NGC 6240
& $14.64\pm0.13$ & $352\pm3$ & $12.2\pm1.1$ & --- & $1.2\pm0.1$ 
& $16.2\pm0.2$ & $447\pm5$ & --- & --- &  --- \\
NGC 7582& $4.65\pm0.04$ & $80.6\pm0.8$ & $0.65\pm0.03$ & $51\pm2$ & $7.1\pm0.3$ 
& $13\pm2$ & $60\pm20$ & $7.1\pm1.5$ & $-120\pm5$ & $120\pm2$ \\
\enddata
\tablecomments{1) Target name (2) EW of the [\ion{Fe}{2}] line. The upper limit of three times of the error (standard deviation) is indicated for no detection. (3) The line broadening of [\ion{Fe}{2}], which is shown only when the line is modeled by a single Gaussian. (4) and (5), (7) and (8), (9) and (11) are the same as (2) and (3), but for the [\ion{P}{2}] line, the central peak of [\ion{S}{3}], and the blue wing of [\ion{S}{3}], respectively. (10) Velocity shift for the  line center of [\ion{S}{3}] blue wing from the central peak.}
\end{splitdeluxetable*}

\subsection{The [\ion{S}{3}] line}

\begin{figure*}
\gridline{\rotatefig{0}{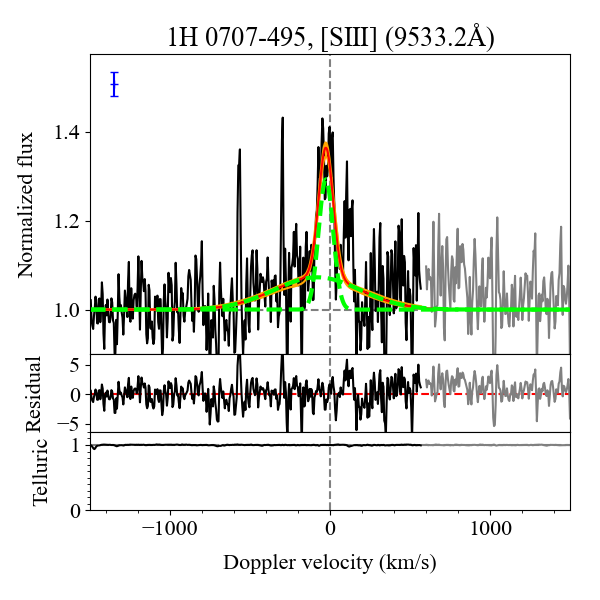}{0.25\textwidth}{(a)}
          \rotatefig{0}{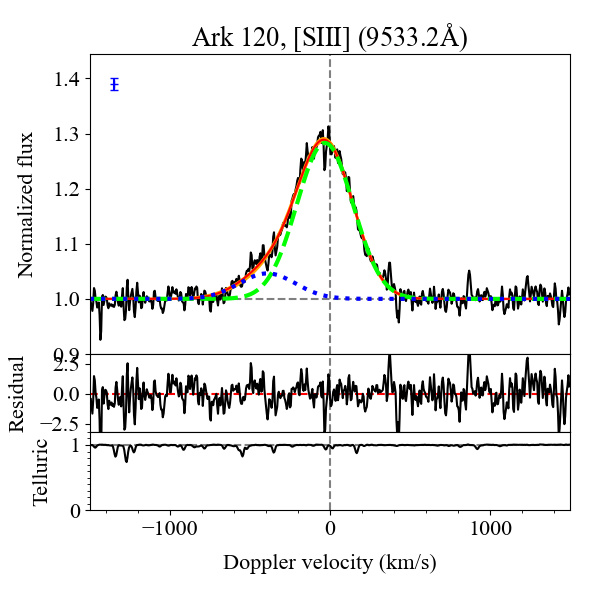}{0.25\textwidth}{(b)}
          \rotatefig{0}{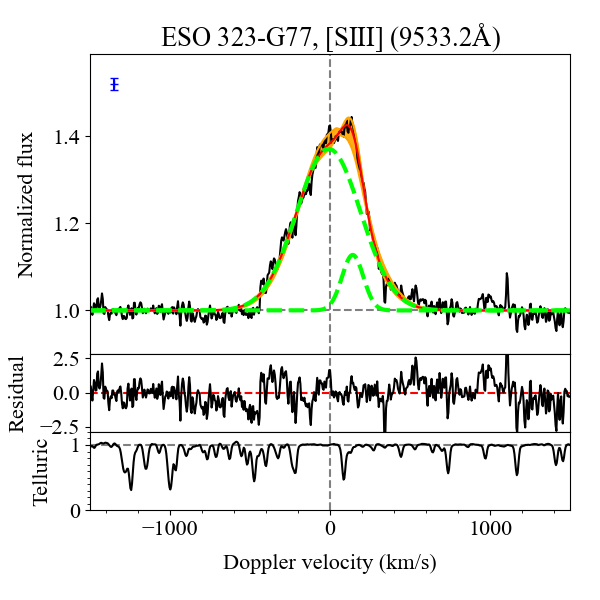}{0.25\textwidth}{(c)}
           \rotatefig{0}{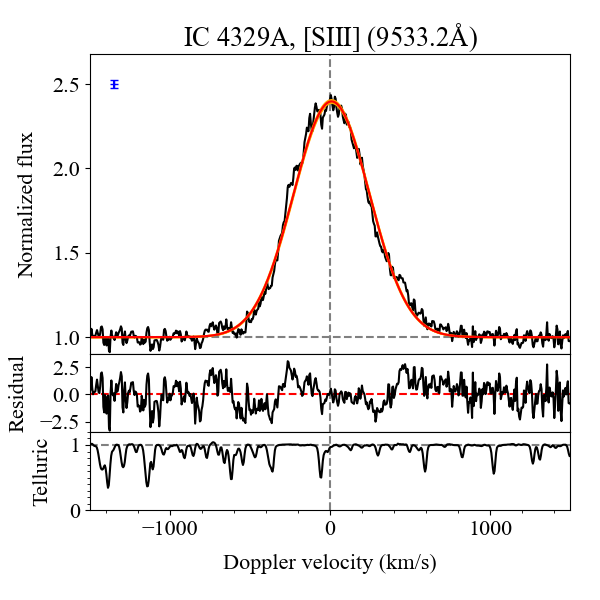}{0.25\textwidth}{(d)}
          }
\gridline{
           \rotatefig{0}{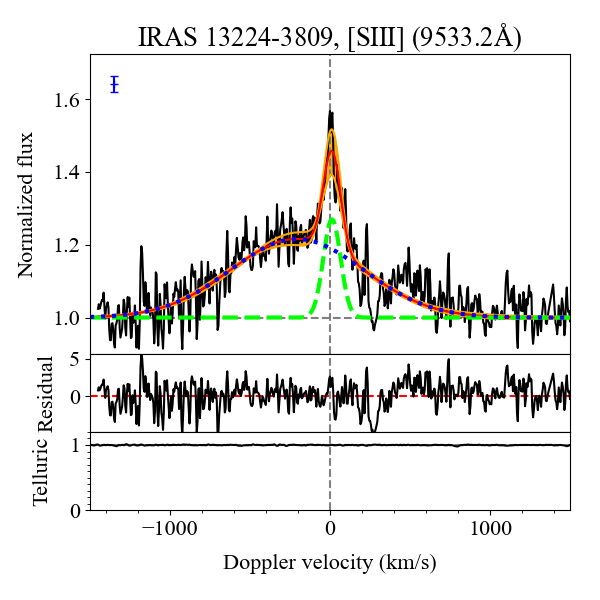}{0.25\textwidth}{(e)}
           \rotatefig{0}{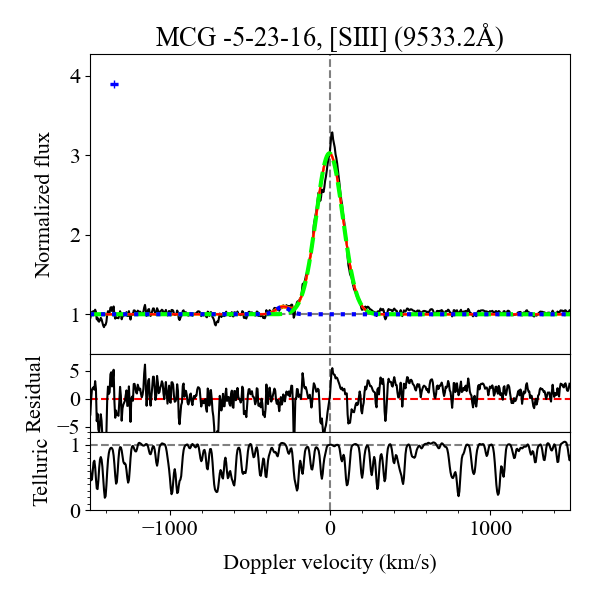}{0.25\textwidth}{(f)}
           \rotatefig{0}{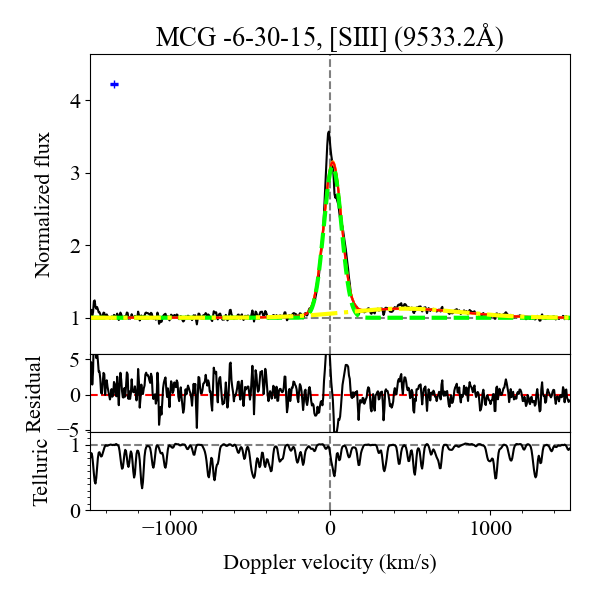}{0.25\textwidth}{(g)}
           \rotatefig{0}{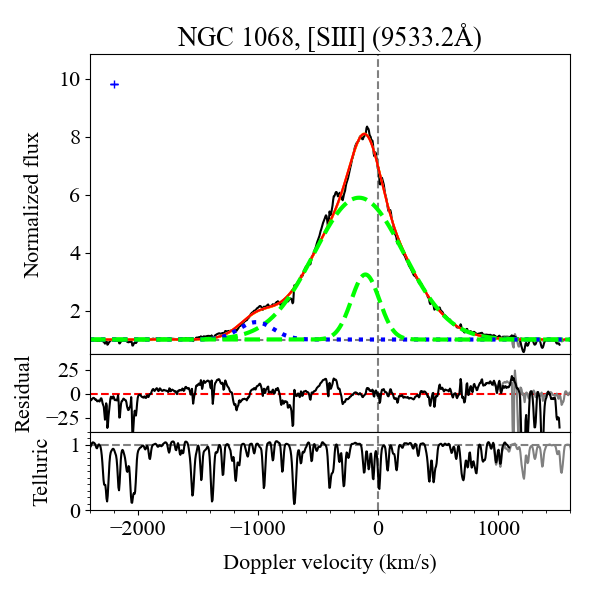}{0.25\textwidth}{(h)}
} 
\gridline{           
\rotatefig{0}{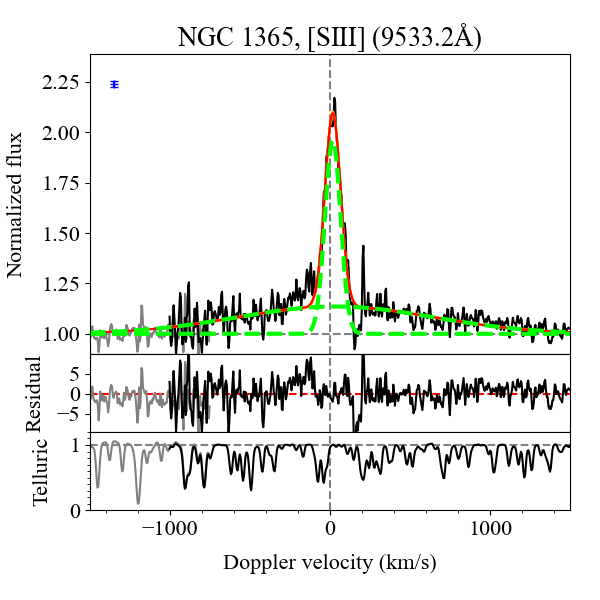}{0.25\textwidth}{(i)}
\rotatefig{0}{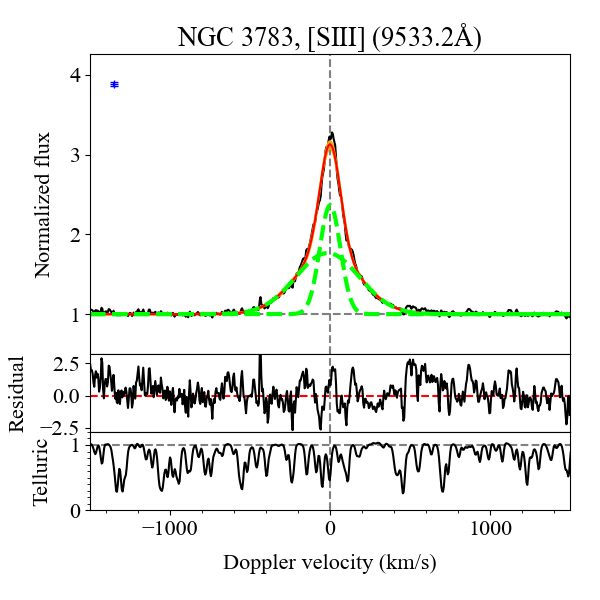}{0.25\textwidth}{(j)}
           \rotatefig{0}{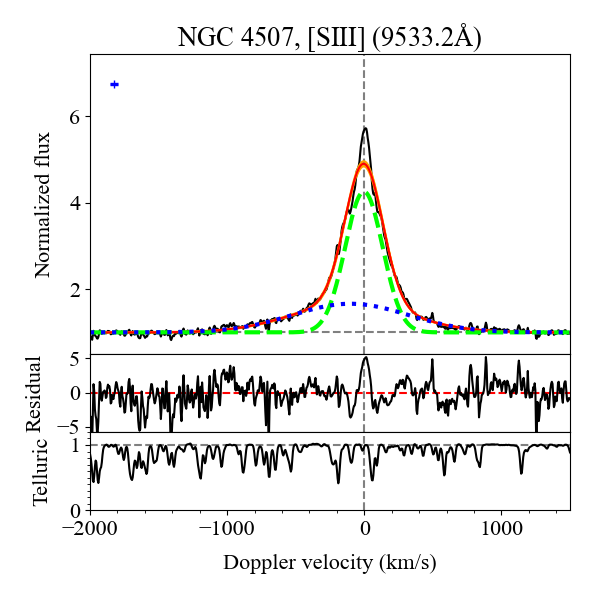}{0.25\textwidth}{(k)}
           \rotatefig{0}{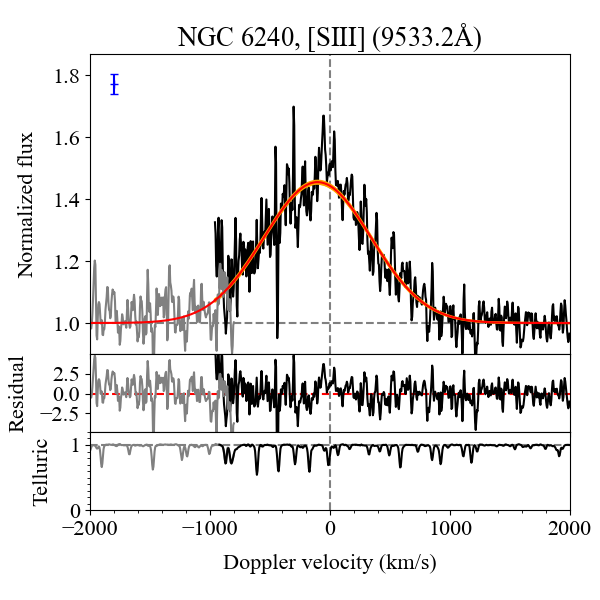}{0.25\textwidth}{(l)}
           }    
\gridline{\leftfig{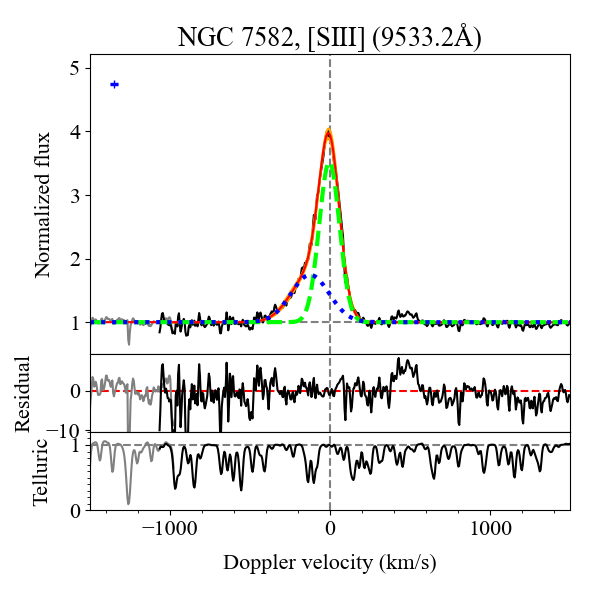}{0.25\textwidth}{(m)}
          }
          \caption{Velocity profiles of the [\ion{S}{3}] emission lines.
          The gray solid line shows the spectrum in the adjacent Echelle order.
          Gaussians whose central velocity is $v=0$ within $3\sigma$ are shown in the green dashed lines, whereas those on the blue side ("blue wing") are in the blue dotted lines. The exceptional case is MCG $-$6-30-15, which shows an excess on the red side (yellow dot-dashed line).
          The other notations are the same as those in Figure \ref{fig:lineprofile}.}
\label{fig:SIII}
\end{figure*}

We use the [\ion{S}{3}]$\lambda9533$ line as a tracer of the ionized outflow, instead of the conventional optical [\ion{O}{3}]$\lambda5007$ line.
The two lines share the same transition ($^3$P$_2$--$^1$D$_2$) and belong to the same family in the periodic table.
It is  predicted that they are emitted from the similar region in the AGN (see e.g., \citealt{liu15}).
We also note that sulfur is an almost perfect volatile element since it is not locked in dust and thus is suitable for tracing the pure outflow component. 
Figure \ref{fig:SIII} shows the obtained [\ion{S}{3}]$\lambda9533$ line profiles.
The fitting procedure is almost the same as the one in Figure \ref{fig:lineprofile}, but the line center is allowed to be $\pm1000$~km~s$^{-1}$ because the gas motion traced by the [\ion{S}{3}] may be larger. 
The first Gaussian to be introduced, whose line velocity is almost zero in the rest frame, is referred to as the ``main'' one, and the additional ones are as ``sub''.
If the center velocity of the ``sub'' Gaussian was negative (within a range of $-1000$~km~s$^{-1}$ to 0~km~s$^{-1}$) at a 99.7\% confidence level, it was interpreted as the ``blue wing'' and we categorized that the object has the ``ionized outflow''.
Consequently, in 6 out of the 13 AGNs (Ark 120, IRAS 13224$-$3809, MCG $-$5-23-16, NGC 1068, NGC 4507, NGC 7582),  the ``blue wing'' is detected.
Table \ref{tab:result} lists the fitting results, including the velocity shifts, $\sigma_v$, and EWs.

\subsection{Absorption features around the [\ion{P}{2}] line} \label{sec:otherabs}
Some absorption features  were  apparent in some AGNs around the [\ion{P}{2}] line.
 In particular, those at $v=-1400$~km~s$^{-1}$ and $+200$~km~s$^{-1}$ (corresponding to 11830~\AA\ and 11895~\AA, respectively),  were  observed in multiple sources:
the 11830~\AA\ line  in MCG $-$5-23-16, MCG $-$6-30-15, NGC 1068, NGC 1365, NGC 4507, and NGC 7582,  and the 11895~\AA\ line in ESO 323-G77, NGC 1365, and NGC 7582.
These lines cannot be the remaining counts of the telluric absorption because some of them are seen in the wavelength range in which any potential remaining effect is small.
They were broader than the telluric lines and instrumental resolution. Therefore they are not artificial.
They  were located at a similar wavelength in the rest frame of each AGN, implying that they are associated with AGNs and not the Galactic absorption.
Interestingly, they  were   present mostly in Seyfert 2 galaxies; the exceptional cases are MCG $-$6-30-15 (which has complex absorbers close to the central SMBH; \citealt{lee01}) and NGC 1365 (whose broad H$\alpha$ line is extremely faint and classified as Seyfert 1.9 \citealt{tri10}. This implies that most of the broad line region may be covered).
 Consequently, we suggest that these absorptions are due to molecular gas in the torus or some obscuring material (e.g., polar dust; \citealt{honig2013}) in the vicinity of AGNs.
The line identification and search for other features will be  presented in  our forthcoming paper.

\subsection{Spatial distribution of the [\ion{Fe}{2}]/[\ion{P}{2}] ratio in NGC 1068} \label{sec:NGC1068spatial}
\begin{figure}
    \centering
    \includegraphics[width=5.5cm,angle=270]{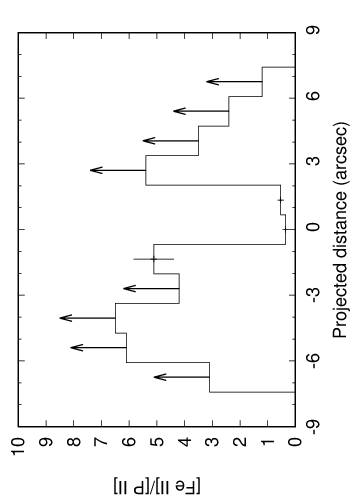}
    \caption{The [\ion{Fe}{2}]/[\ion{P}{2}] ratio in NGC 1068 as a function of projected distance from the core, i.e., the continuum peak. The positive direction was determined to be northwest.}
    \label{fig:longslit}
\end{figure}

NGC 1068 is the brightest and most nearby targets and thus suitable for studying  position dependence on the [\ion{Fe}{2}]/[\ion{P}{2}] ratio;
it was reported that, whereas the [\ion{Fe}{2}]/[\ion{P}{2}] ratio is small ($\sim1$) near the central core, it increases up to $\sim7$ at a distance of 2--4 arcsec from the center \citep{has11,rif14}.
In order to confirm such spatial dependence in our observation, we created spectra divided in the slit-length direction and calculated the EWs of the two emission lines.
The continuum peak position was set as the origin, and the spectra were extracted from every 5 pixels (= 1.35 arcsec; almost equivalent to the seeing).
The position angle was 220~deg and the slit was oriented roughly from southeast to northwest.
Eleven spectra were derived, covering 7.5 arcsec on each side in total. 
Whereas the [\ion{Fe}{2}] emission lines were detected in all of them, the [\ion{P}{2}] emission lines were detected only in the three central ones.
The EW values or their upper limits were calculated in the same way as in section \ref{sec3.1}. The width of the undetected [\ion{P}{2}] lines used for the upper limit was the same as that for the spectrum at the central bin.
The obtained spatial dependence of the [\ion{Fe}{2}]/[\ion{P}{2}] ratio is shown in Figure \ref{fig:longslit}. 
As in the previous studies \citep{has11,rif14}, we confirmed that the ratio is smaller at the center and higher at a projected distance of $\sim3$ arcsec.
This means that photo-ionization is dominant in the vicinity of the core in NGC 1068, while the shock ionization appears farther from the core.

\section{Discussion}\label{sec4}

\subsection{The [\ion{Fe}{2}]/[\ion{P}{2}] ratio}
\begin{figure}
    \centering
    \includegraphics[width=8cm]{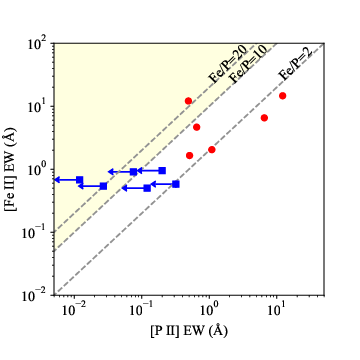}
    \caption{Relation between the EWs of [\ion{Fe}{2}] and [\ion{P}{2}].
    Red points show the targets  from which both  the [\ion{Fe}{2}] and [\ion{P}{2}] lines are detected, whereas  blue points do those with no detection of  [\ion{P}{2}]. Dashed lines show [\ion{Fe}{2}]/[\ion{P}{2}]=2, 10, and 20.
    The yellow-shaded area shows the shock region.}
    \label{fig:P_Fe}
\end{figure}

Figure \ref{fig:P_Fe} shows the relation between the EWs of the two emission lines, [\ion{Fe}{2}] and [\ion{P}{2}]. 
All the sources for which [\ion{P}{2}]  is not detected show only a weak [\ion{Fe}{2}] line.
This suggests that non-detection of [\ion{P}{2}] is simply because the narrow lines are weak regardless of the species, rather than due to the existence of shock.
The detection of weak [\ion{P}{2}] emission lines is critical for accurate measurement of the [\ion{Fe}{2}]/[\ion{P}{2}] ratio.
For example, the weakest [\ion{P}{2}] line detected in T16 (NGC 5506) has an EW of $\sim1.8$~\AA\ (n.b.\ this value is not very accurate because it is read from the spectrum figure).
On the other hand, our study can detect the emission line with EW = 0.49~\AA\ (NGC 4507), which is 3.6 times fainter than the detection limit of T16.
For example, if the detection limit of the [\ion{P}{2}] line were 1.8~\AA\ in NGC 4507 like T16,
then the [\ion{P}{2}] line could not be detected and the ratio value had only a lower limit of $>$6.7. 
In this case, we could not conclude whether the shock exists.
Consequently, our observation provides a more robust conclusion on the existence of shock than the previous study.

\begin{figure}
    \centering
    \includegraphics[width=8cm]{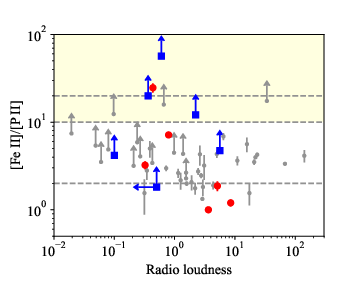}
    \caption{Relation between the [\ion{Fe}{2}]/[\ion{P}{2}] ratio and radio loudness.
    The notations are the same as in Figure \ref{fig:P_Fe}. The results of T16 are  superposed in gray. We find no correlation between the existence of shocks and radio loudness.}
    \label{fig:R_ratio}
\end{figure}

Figure \ref{fig:R_ratio} compares the [\ion{Fe}{2}]/[\ion{P}{2}] ratios with the radio loudness. The result of T16 is also  superposed.
T16 found that no significant difference exists between the average value of the [\ion{Fe}{2}]/[\ion{P}{2}] ratios in the radio-loud  AGNs ($R>10$; the average value is $3.67\pm1.09$) and that in the radio-quiet AGNs ($R<10$; the average value is $2.87\pm1.38$) for the [\ion{P}{2}]-detected objects,
and stated that the radio jet  is unlikely  to be the main contributor to the shock in the NLR. 
We performed a Kolmogorov-Smirnov (KS) test to quantitatively compare the distributions of the [\ion{Fe}{2}]/[\ion{P}{2}] ratios between this study and T16.
Only the [\ion{P}{2}]-detected sources were used.
The obtained $p$-value was 0.562, so that 
We cannot discuss any significant differences in this sample.

\begin{figure}
    \centering
    \includegraphics[width=8cm]{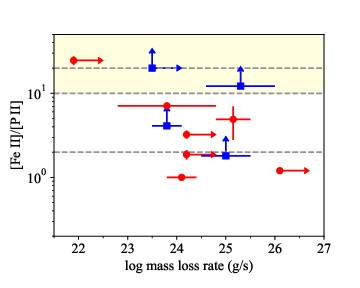}
    \includegraphics[width=8cm]{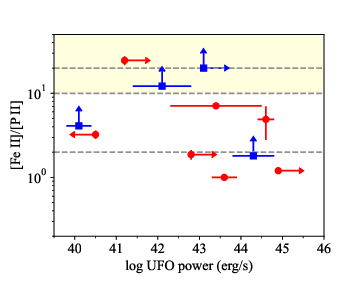}
    \caption{Relation between the [\ion{Fe}{2}]/[\ion{P}{2}] ratio and  UFO mass-loss rate (in\ upper\ panel) and kinetic power (in\ lower\ panel). The notations are the  same as in Figure \ref{fig:P_Fe}. No correlation between the existence of shock and the UFO parameter.
    }
    \label{fig:K_ratio}
\end{figure}

\subsection{UFO contribution}
To study the UFO contribution to the shock, 
we compare the UFO mass loss rate and kinetic power with the [\ion{Fe}{2}]/[\ion{P}{2}] ratios in Figure \ref{fig:K_ratio}.
Using the four objects for which both UFO parameters and the [\ion{Fe}{2}]/[\ion{P}{2}] ratios were determined (NGC 7582, NGC 1068, IRAS 13224--3809, NGC 6240), we examined their correlations. The Pearson correlation coefficient values and $p$-values were --0.465 and 0.535 for log mass loss rate, and --0.368 and 0.632 for log UFO power, respectively. Thus, no correlations were detected between them.

While the UFO is considered to sweep through the ISM and produce a kpc-scale outflow (e.g., \citealt{tom15,mar20}),
the efficiency of the kinetic power transfer from the UFO to the kpc-scale outflow has a wide variety, depending on the flow type; notably, it can be in the energy conserving or momentum conserving modes (see, e.g., \citealt{miz19}).
If these two types exist in the sample, no correlation is observed even if UFOs have some (indirect) contribution to kpc-scale outflows and shocks.
There is a possibility that if only AGNs with energy-conserving flows were selected, a correlation between UFOs and NLR-scale shocks  might be identified.  
Our  sample size is, however, too  small  to conduct this type of study.
Determination of the flow type from sub-pc to kpc will be required for a direct assessment of the  effect of UFOs to shocks.

\subsection{Correlation with the [\ion{S}{3}] outflow}

\begin{figure}
    \centering
    \includegraphics[width=8cm]{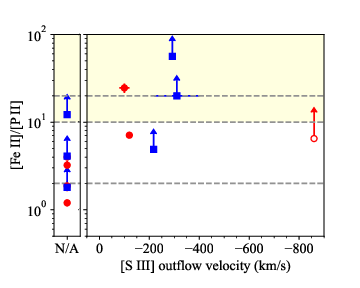}
    \caption{ Relation between the [\ion{Fe}{2}]/[\ion{P}{2}] ratio and the [\ion{S}{3}] outflow velocity. The outflow velocity of the targets with no [\ion{S}{3}] outflow is set  to 0. 
    The open red circle is for NGC 1068 but off-centered (where the projected distance is $\sim-3$~arcsec; see Figure \ref{fig:longslit}).
    The other notations are the same as in Figure \ref{fig:P_Fe}.
    }
    \label{fig:s3_ratio}
\end{figure}

Figure \ref{fig:s3_ratio} shows the obtained relation between the [\ion{S}{3}] outflow velocity, that is, the velocity shift in the center of the blue wing from the rest frame if the source has an outflow, and the line ratio. 
We use the off-centered value for NGC 1068.
There is a hint that targets associated with  [\ion{S}{3}] outflows have a larger [\ion{Fe}{2}]/[\ion{P}{2}] ratio, that is, they are more likely to produce shocks.

As discussed in section \ref{sec:NGC1068spatial},
the [\ion{Fe}{2}]/[\ion{P}{2}] ratio of NGC 1068 is 
higher ($\sim$7) at an angular distance of $\sim3$~arcsec,
whereas small ($<1$) within 1~arcsec from the center.
In addition, integral field unit (IFU) observations revealed an annulus of $r\sim2$~arcsec with high [\ion{Fe}{2}]/[\ion{P}{2}] ratio ($\sim$5--8), whereas an inner circular region of $r<1$~arcsec has a small ratio ($\lesssim2$).
This kind of off-centered, high [\ion{Fe}{2}]/[\ion{P}{2}] regions have been also found in other AGNs (e.g., \citealt{sto09,sch19,rif20}).
If we use the [\ion{Fe}{2}]/[\ion{P}{2}] ratio in the off-centered region (see Figure \ref{fig:longslit}),
the ratio is similar to the ones in the other objects with the [\ion{S}{3}] outflows (see the open red circle in Figure \ref{fig:s3_ratio}).
Consequently, all the AGNs with the NLR-scale outflows are found to be high [\ion{Fe}{2}]/[\ion{P}{2}] ratio of $\gtrsim7$ in our sample, and thus the ionized outflows are considered to play a dominant role to the shock ionization.

\begin{figure}
    \centering
    \includegraphics[width=8cm]{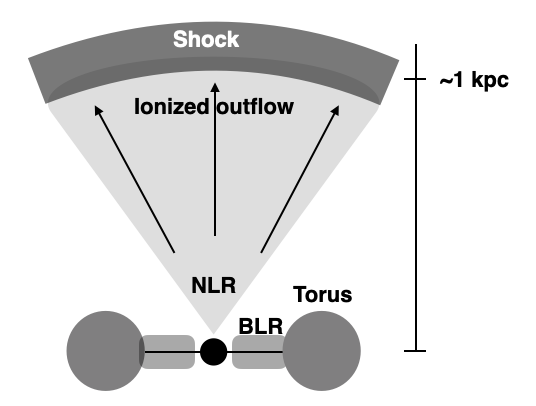}
    \caption{Schematic picture of the ionized outflow, NLR, and shock region. In the NGC 1068 case, the ionized outflow is traced in the [\ion{O}{3}] and [\ion{S}{3}] lines. The outer radius of the ionized outflow matches the radius of the shock-dominant annulus.}
    \label{fig:schematic}
\end{figure}

In addition, we notice that the [\ion{Fe}{2}]/[\ion{P}{2}] peak region in NGC 1068 reported by \citet{rif14} and the [\ion{O}{3}] spatial distribution shown by \citet{das06} share a similar size scale.
The [\ion{O}{3}] emission region is extended toward the northeast direction with a radius of 0--2~arcsec \citep{das06}.
This means that the ionized outflow is blown from the AGN core in this direction.
On the other hand, the annulus with high [\ion{Fe}{2}]/[\ion{P}{2}] ratio ($r\sim2$--3~arcsec) has asymmetry and the shock is more prominent in the northeast part \citep{rif14}.
These geometry matches our scenario that the AGN ionized outflow triggers the shock ionization, shown in Figure \ref{fig:schematic}.
Based on this picture, the kinetic energy or the momentum flux of the ionized outflow triggered by AGN are transferred to the gas in a several-hundred-pc-scale region via the shock.
In other words, the AGN  feedback (i.e., AGN affects to the host galaxy via the outflow) occurs in the NLR via the ionized outflow and the shock.

The exception is  ESO 323-G77, which is not associated with the [\ion{S}{3}] outflows but has a ratio larger than ten. 
The [\ion{S}{3}] emission line of ESO 323-G77 shows an asymmetric profile,
which is present only  in this source among our sample (Figure \ref{fig:SIII}).
If  the asymmetry  originates in the outward motion of the NLR cloud in a dusty region \citep{peterson}, such a bulk motion may be responsible for some shock ionization.
 Another possible scenario is that a potential dusty wind, the presence of which was suggested based on the detection of dust emission  extending axially in this source \citep{lef18}, generates  the shocks.
 In any case, some outflowing gas may exist in ESO 323-G77.

\subsection{Future work}
In our observations, some AGNs exhibit [\ion{P}{2}] lines too weak to be detected, necessitating the calculation of upper limits on the [\ion{Fe}{2}]/[\ion{P}{2}] ratios only. In these cases, firm conclusions regarding the presence of shocks for these targets are elusive. Therefore, it is neccesary to employ multiple diagnostic methods to confirm the existence of shocks. For instance, Figure 13 in \citet{calabro2023} introduces various diagnostic diagrams for shock identification. Among these, the [\ion{S}{3}] 9530\AA /Pa$\gamma$ vs [\ion{Fe}{2}] 1.257~$\mathrm{\mu m}$/Pa$\beta$ diagram can be constructed based on our WINERED observations. Our subsequent publications will provide comprehensive spectral data, including full spectra, line profiles, and equivalent widths (EWs), encompassing Pa$\beta$ and Pa$\gamma$. 

Spatially-resolved observations of [\ion{Fe}{2}], [\ion{P}{2}], and [\ion{S}{3}] play a pivotal role in elucidating the ionization mechanism within the NLR and facilitating direct observations of energy transfer from the AGN to kpc-scale regions. Some prior studies have generated [\ion{Fe}{2}] and [\ion{P}{2}] maps within the NLR context \citep{sto09,rif14,fazeli2019,sch19,rif20,rif21}. For instance, \citet{sch19} presented [\ion{Fe}{2}]/[\ion{P}{2}] maps for six nearby AGNs, highlighting that NGC 3227 and NGC 5899 exhibit peak values offset from the nuclear position, aligning with the high velocity dispersion of [\ion{Fe}{2}]. However, the sample size remains limited, underscoring the significance of our study in expanding the roster of suitable candidates for such observations.

The schematic diagram depicted in Figure \ref{fig:schematic} closely aligns with theoretical shock models \citep{kin10,zub12,fau12,fau12b,kin15,richings2018,riching2018b,zub19}. In these models, the inner radiative-driven wind interacts with the ISM, giving rise to a shock front. This shock front subsequently sweeps up the ISM, generating a region of shocked ISM. The nature of this shock hinges on whether cooling, typically radiation, significantly dissipates energy from the hot shocked gas on a timescale shorter than its flow time. If cooling exerts a strong influence in this context, most of the pre-shock kinetic energy is lost via radiation, characterizing a momentum-driven flow. Intense cooling results in substantial compression of the shocked gas, yielding a geometrically narrow post-shock region. Conversely, in scenarios where cooling is negligible, the post-shock gas retains all the mechanical energy imparted by the shock and undergoes adiabatic expansion into the ISM. In contrast to the momentum-driven (isothermal) case, the post-shock gas in energy-driven flows is characterized by a geometrically elongated structure. Given that energy-driven flows are significantly more intense than momentum-driven flows, the assessment of radiative outflow's impact on the surrounding environment (in this case, the NLR) hinges on the type of gas flow described above. To unravel these complexities, it is essential to measure momentum fluxes before and after the shock, necessitating an accurate assessment of spatial scales. Therefore, spatially-resolved observations, as outlined in the preceding paragraph, assume paramount importance.

\section{Conclusion}\label{sec5}
 We  performed NIR slit spectroscopy 
 of 13 Seyfert galaxies  for which UFOs had been observed in X-rays, and measured their [\ion{Fe}{2}]/[\ion{P}{2}] ratio, with the aim of determining whether AGN outflows can generate shocks in the NLR.
The [\ion{Fe}{2}] and [\ion{P}{2}] lines  were detected in 12 and 6 sources, respectively, among them.
From the derived ratios, we found that shocks are likely to be present in 4 objects, Ark 120, ESO 323-G77, MCG $-$5-23-16, and NGC 4507.
No direct relation between the presence/absence of the shocks and UFOs is found.
On the other hand, we found that when the [\ion{Fe}{2}]/[\ion{P}{2}] ratio is higher than $\sim7$, the [\ion{S}{3}] emission line shows signs of the ionized outflow. 
We favor a scenario that the ionized outflows trigger the NLR-scale shock.
The scenario  had been  evaluated so far only for NGC 1068, using the IFU observation, before this work.
This  work  revealed that  the scenario is likely to be  applicable for a wider range of sample.

\acknowledgments
This study is based on observations collected at the European Southern Observatory under ESO programme 0100.B-0798(A).
We are deeply grateful to the NTT and La Silla team in ESO for their technical support.
This study was financially supported by the Hayakawa Fund in Astronomical Society of Japan (M.M.), and Grants-in-Aid, KAKENHI, from Japan Society for the Promotion of Science (JSPS) Nos.\ JP21K13958 (M.M.) and JP19K03917 (H.S.).
The WINERED was developed by the University of Tokyo and the Laboratory of
Infrared High-resolution spectroscopy (LiH), Kyoto Sangyo University
under the financial supports of JSPS KAKENHI Nos.\ JP16684001 (N.K.), JP20340042 (N.K.), JP21840052 (Y.I.), and the MEXT Supported Program for the Strategic Research Foundation at Private Universities, Nos.\ S0801061 and S1411028 (H.K.).
N.K. appreciates the support from JSPS-DST under the Japan-India Science Cooperative Programs during 2013--2015 and 2016--2018.



\facility{NTT}
\software{WINERED pipeline (Hamano et al.\ in prep.), IRAF \citep{iraf1,iraf2}, PyRAF \citep{pyraf}
}

\clearpage
\appendix
\section{Line profiles  of each object}\label{sec:each}
 This appendix describes the line profiles of the [\ion{Fe}{2}], [\ion{P}{2}], and [\ion{S}{3}] lines, and other emission/absorption-like features, if any, of individual sources (Figures \ref{fig:lineprofile} and \ref{fig:SIII}).

\subsection{1H 0707$-$495}
The data of 1H 0707$-$495 has  the smallest signal-to-noise ratio  among the  observed target sources, $S/N\sim15\textendash20$ at the continuum level.
The [\ion{Fe}{2}] emission line was detected and fitted by a single Gaussian, while not for [\ion{P}{2}].
Some residuals are seen around $-1400$~km~s$^{-1}$ of the [\ion{P}{2}] line, the origin of which is unknown.
The [\ion{S}{3}] line has a narrow core ($\sigma_v=46\pm3$~km~s$^{-1}$) and broad emission ($\sigma_v=280\pm30$~km~s$^{-1}$).

\subsection{Ark 120}
At the [\ion{P}{2}] wavelengths no emission lines are detected, but rather a slight absorption structure can be seen.
The [\ion{S}{3}] blue wing is  present at $-310\pm90$~km~s$^{-1}$.

\subsection{ESO 323-G77}
The [\ion{S}{3}] line  shows an asymmetric profile and is well fitted with two Gaussians; the main one has $\sigma_v=203\pm4$~km~s$^{-1}$ and the sub one has the line shift of $+144\pm9$~km~s$^{-1}$ and $\sigma_v=75\pm11$~km~s$^{-1}$.

\subsection{IC 4329A}
The [\ion{S}{9}]$\lambda12520$ line is  present in addition to the standard lines  and an extra Gaussian is requited to fit the [\ion{Fe}{2}] line properly.

\subsection{IRAS 13224$-$3809}
The data of both  the [\ion{Fe}{2}] and [\ion{P}{2}] lines  appear noisy due to a low signal-to-noise ratio and heavy telluric absorption.
A hint of [\ion{P}{2}] emission line is seen at $v=-100$~km~s$^{-1}$, but its strength ($2.78\sigma$) is below the detection threshold.

\subsection{MCG $-$5-23-16}
The [\ion{S}{9}]$\lambda12520$ line is  also present in addition to the standard lines.
Three absorption lines are  observed in the [\ion{P}{2}] band: $-1400$~km~s$^{-1}$, $-200$~km~s$^{-1}$, and +200~km~s$^{-1}$.
Their line widths are broader than those in the telluric lines. 
 In particular, the one  at $-1400$~km~s$^{-1}$  (11830~\AA) is the broadest among them. The upper limit of the [\ion{P}{2}] is well constrained.
The [\ion{S}{3}] line shows a slight absorption at the center, or it may be due to  two narrow emission lines constituting the central peak. We fit it  with a single positive Gaussian.

\subsection{MCG $-$6-30-15}
The [\ion{S}{9}]$\lambda12520$ line  appears to have a blue wing. 
Absorption features are  present at $-1400$~km~s$^{-1}$ and +100~km~s$^{-1}$ of the [\ion{P}{2}] line.
The [\ion{S}{3}] line has a redshifted broad emission line, the physical origin of which is yet to be  known.

\subsection{NGC 1068}
NGC 1068 is the nearest among our targets. It  shows the largest [\ion{P}{2}] EW and $\sigma_v$ and  smallest [\ion{Fe}{2}]/[\ion{P}{2}] ratio.
The central peak of [\ion{P}{2}] is shifted with $-200$~km~s$^{-1}$  from that of [\ion{Fe}{2}]. 
 An absorption feature  at $-1400$~km~s$^{-1}$ is also  present in the [\ion{P}{2}] band.
The [\ion{S}{3}] outflow has the fastest velocity among our targets, $-840$~km~s$^{-1}$.

\subsection{NGC 1365}
Neither of the [\ion{Fe}{2}] and [\ion{P}{2}] lines  is detected.
The [\ion{Fe}{2}] band has two absorption features at $-50$~km~s$^{-1}$ and $+200$~km~s$^{-1}$, whereas the [\ion{P}{2}] band, at $-1400$~km~s$^{-1}$ and $+200$~km~s$^{-1}$.
The presence of the $+200$~km~s$^{-1}$ features  may be due to either the Doppler motion of the lines or  simply a positional coincidence.
The [\ion{S}{3}] line was modeled by the two Gaussians; a narrower one ($\sigma_v=49\pm5$~km~s$^{-1}$) and a broader one ($\sigma_v=620\pm40$~km~s$^{-1}$). The broader one is slightly redshifted but within the error bar ($+17\pm18$~km~s$^{-1}$).

\subsection{NGC 3783}
The wavelength range  for the [\ion{Fe}{2}] lines of this source  overlaps with  that with heavy telluric absorption.  Nevertheless,
the  wavelength resolution is sufficiently good to enables us to remove  the effect.
The [\ion{S}{9}]$\lambda12520$ line seems to have a red wing, in contrast to that of MCG $-$6-30-15.
A positive residual remains from +500 to +800~km~s$^{-1}$ of the [\ion{P}{2}] line, whose origin is yet to be known.
The [\ion{S}{3}] line profile is  reproduced with  two positive Gaussians; a narrower one ($\sigma_v=65\pm2$~km~s$^{-1}$) and a broader one ($\sigma_v=200\pm3$~km~s$^{-1}$).

\subsection{NGC 4507}
The blue wing is seen in [\ion{Fe}{2}] and merged with the [\ion{S}{9}] line. Therefore, the fitting requires three Gaussians, the [\ion{Fe}{2}] core ($\sigma_v=151\pm2$~km~s$^{-1}$), the [\ion{Fe}{2}] wing ($v=-481\pm4$~km~s$^{-1}$ and $\sigma_v=583\pm16$~km~s$^{-1}$), and the [\ion{S}{9}] line.
The [\ion{S}{3}] line also has the blue wing at $-100\pm20$~km~s$^{-1}$ with $\sigma_v=420\pm120$~km~s$^{-1}$. 
Considering that the [\ion{P}{2}] line has no hint of blue wing,
some shock due to the outflow is expected in this target.
We also note that a strong absorption is present in the [\ion{P}{2}] band ($-1400$~km~s$^{-1}$).

\subsection{NGC 6240}
NGC 6240 is a merging luminous infrared galaxy and has two cores. In this study, the brighter southern core was observed in the slit.
Both  the [\ion{Fe}{2}] and [\ion{P}{2}] lines are strong and broad.  In particular, the [\ion{P}{2}] line profile required two Gaussian component with $v=400\pm80/-440\pm30$~km~s$^{-1}$ and $\sigma_v=850\pm50/370\pm40$~km~s$^{-1}$, respectively. The [\ion{S}{3}] line has a similar line broadening to the [\ion{Fe}{2}] line.

\subsection{NGC 7582}
The absorption features are  observed at around the [\ion{P}{2}] line at $-1400$~km~s$^{-1}$, $+200$~km~s$^{-1}$, and $+1000$~km~s$^{-1}$.
The [\ion{S}{3}] blue wing is detected at $-120$~km~s$^{-1}$.

\section{Corner plots for the [\ion{Fe}{2}] and [\ion{P}{2}] line fitting}\label{sec:contour}
\begin{figure*}
\gridline{\rotatefig{0}{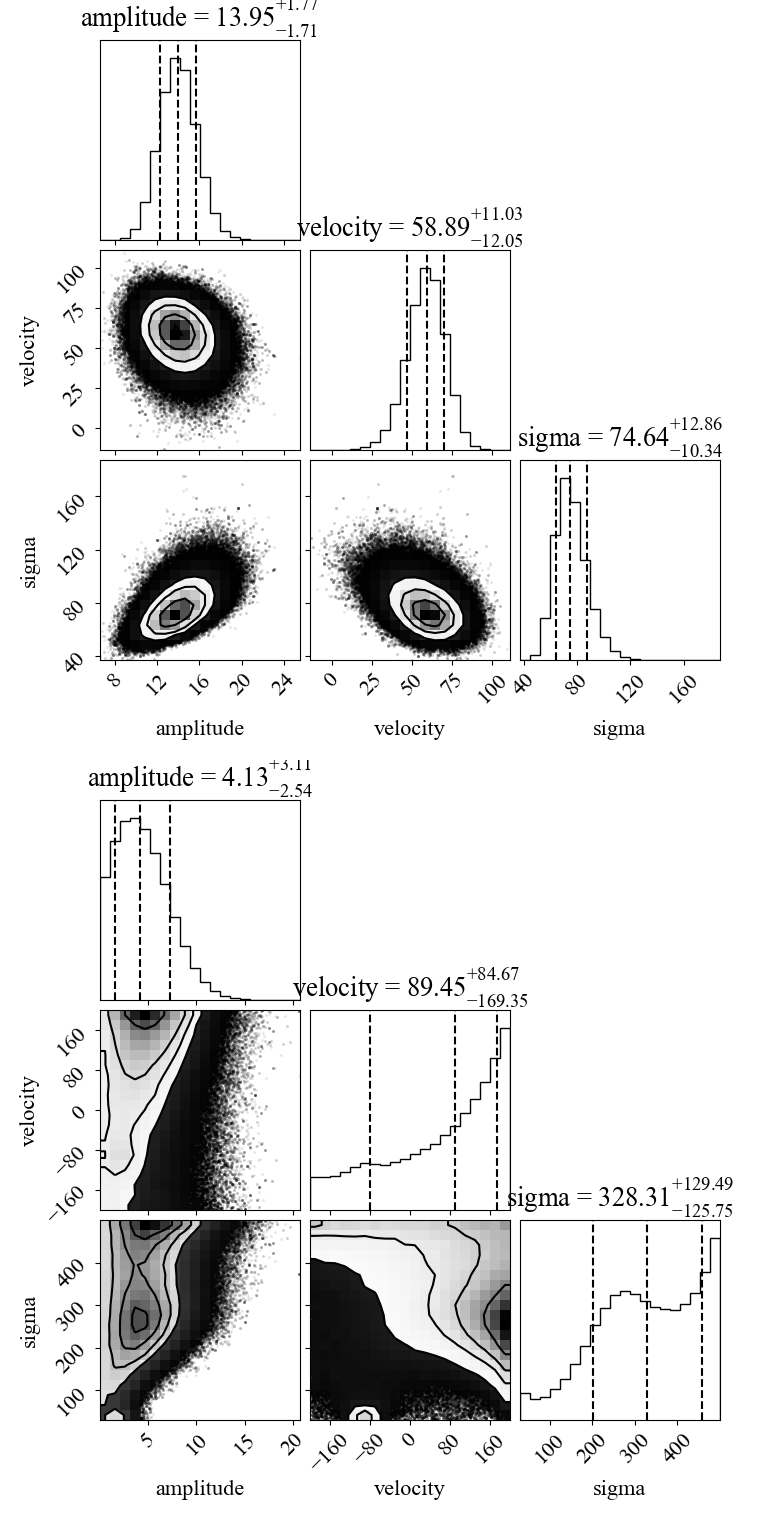}{0.2\textwidth}{(a)}
          \rotatefig{0}{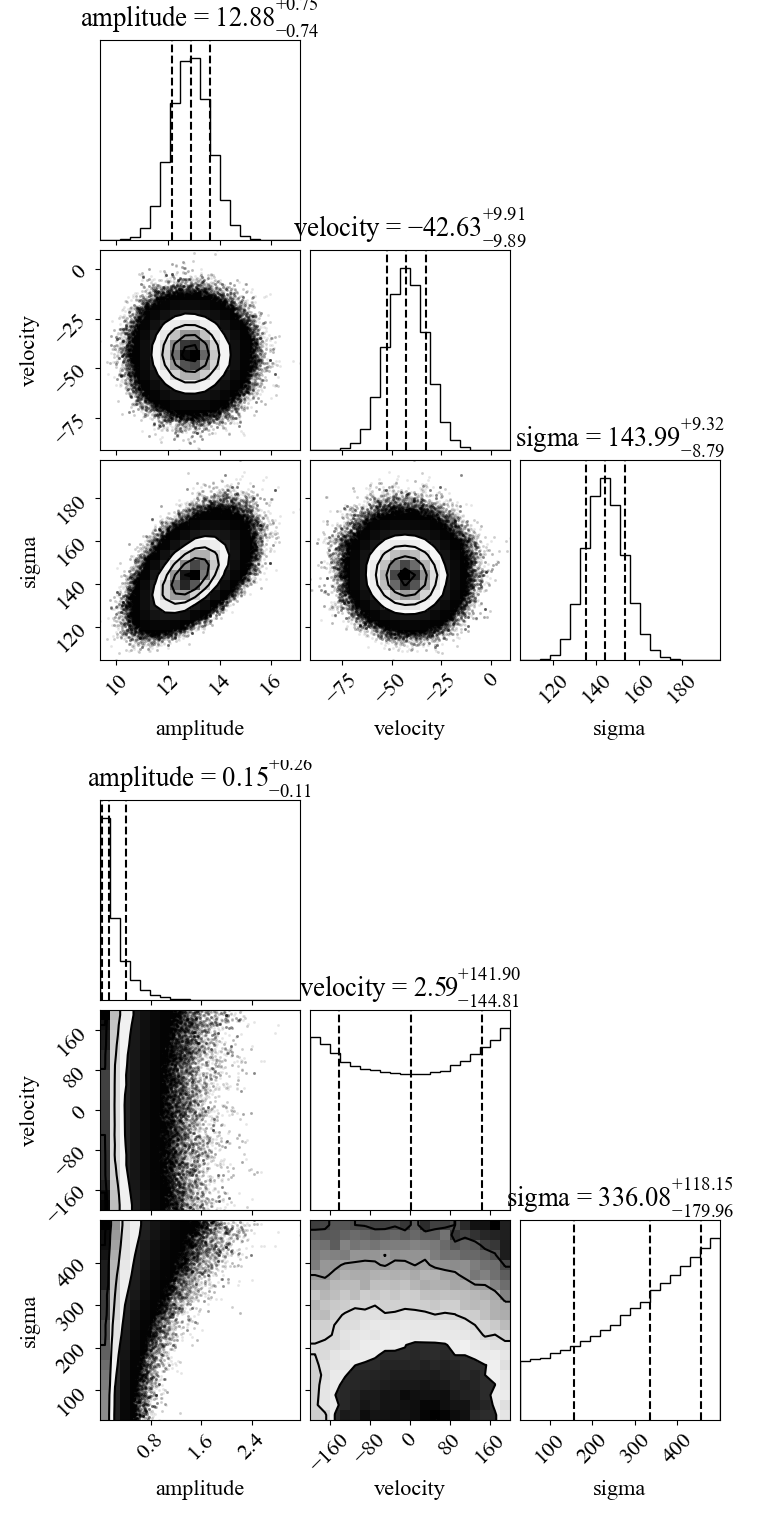}{0.2\textwidth}{(b)}
          \rotatefig{0}{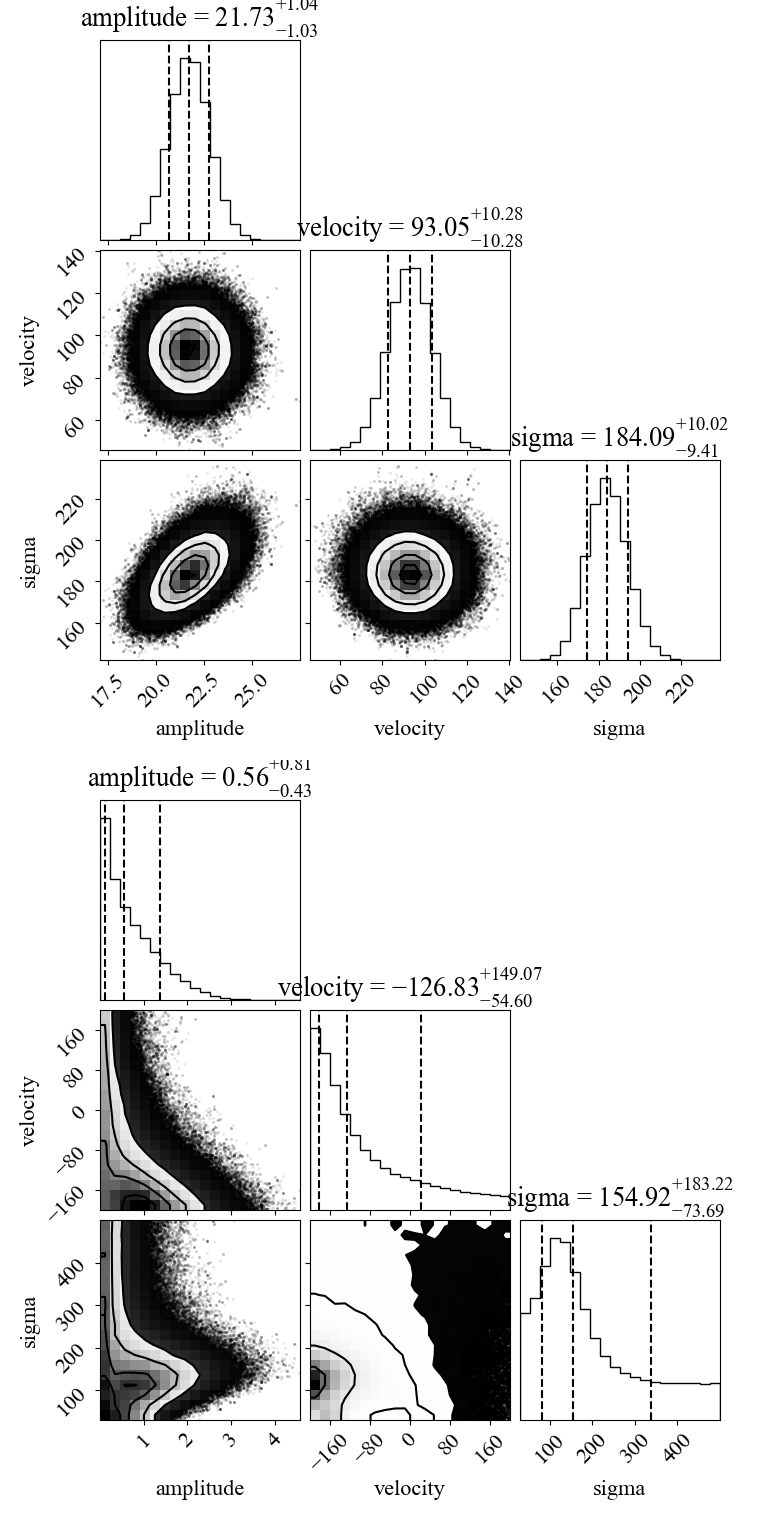}{0.2\textwidth}{(c)}
           \rotatefig{0}{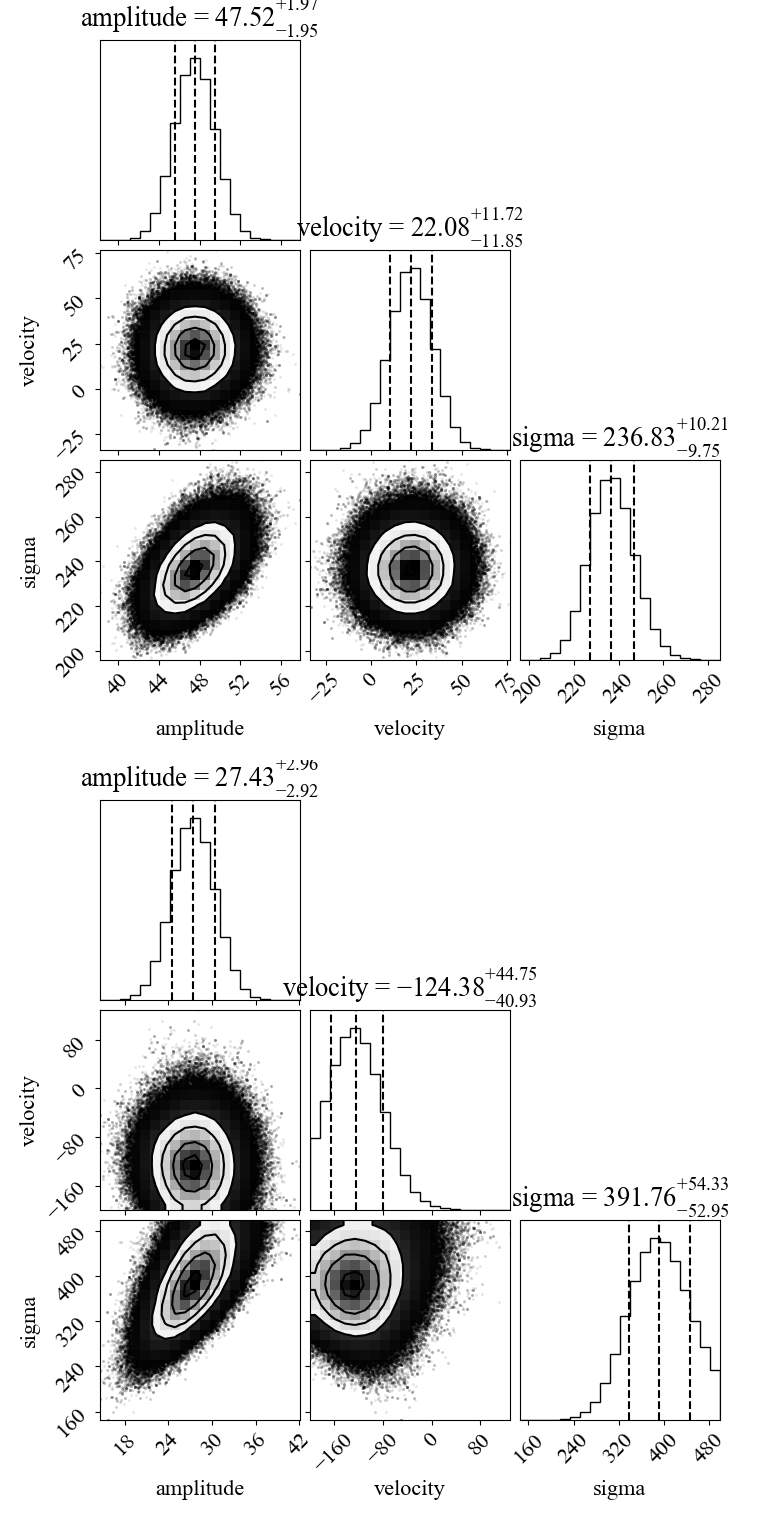}{0.2\textwidth}{(d)}
           \rotatefig{0}{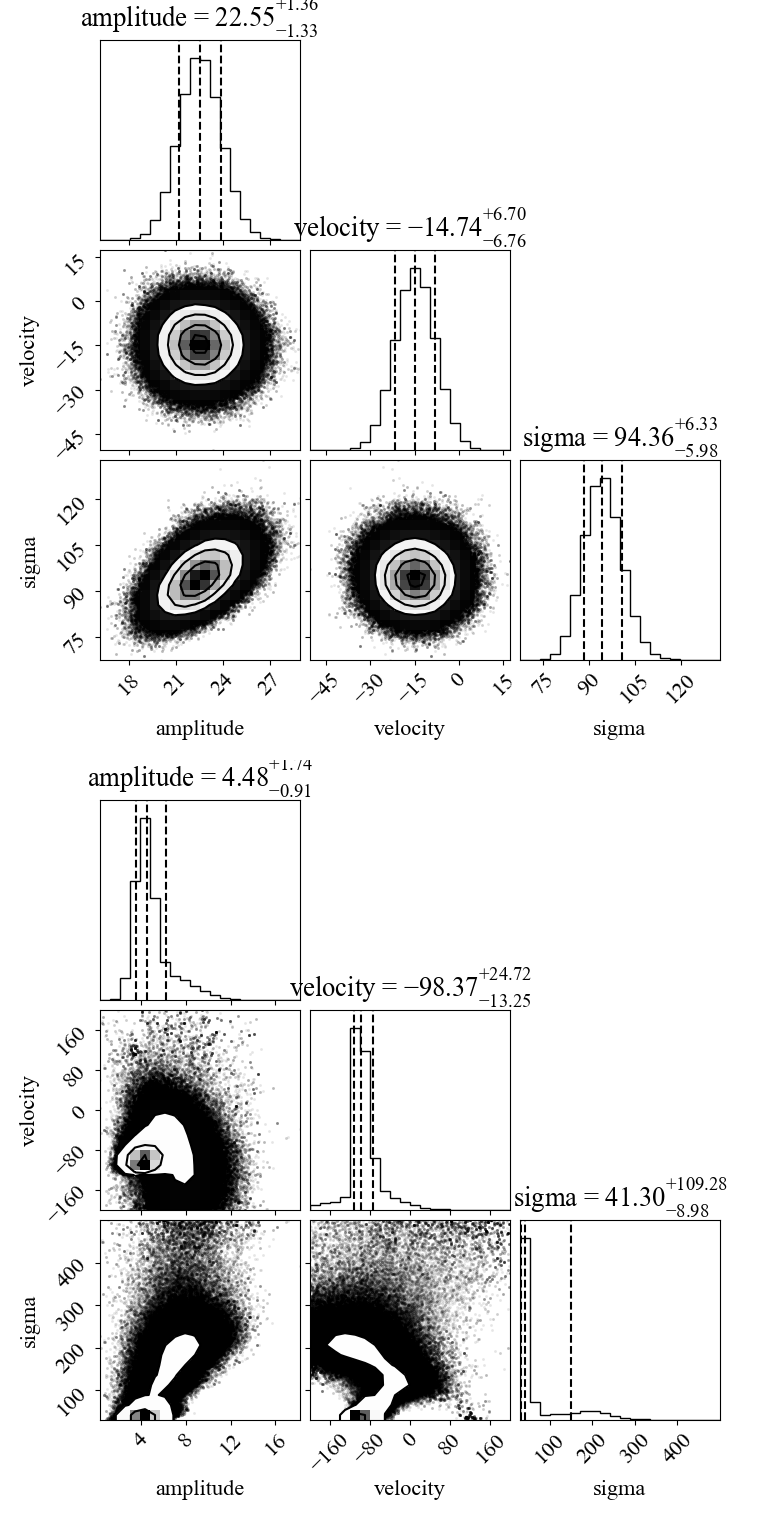}{0.2\textwidth}{(e)}
          }
\gridline{
           \rotatefig{0}{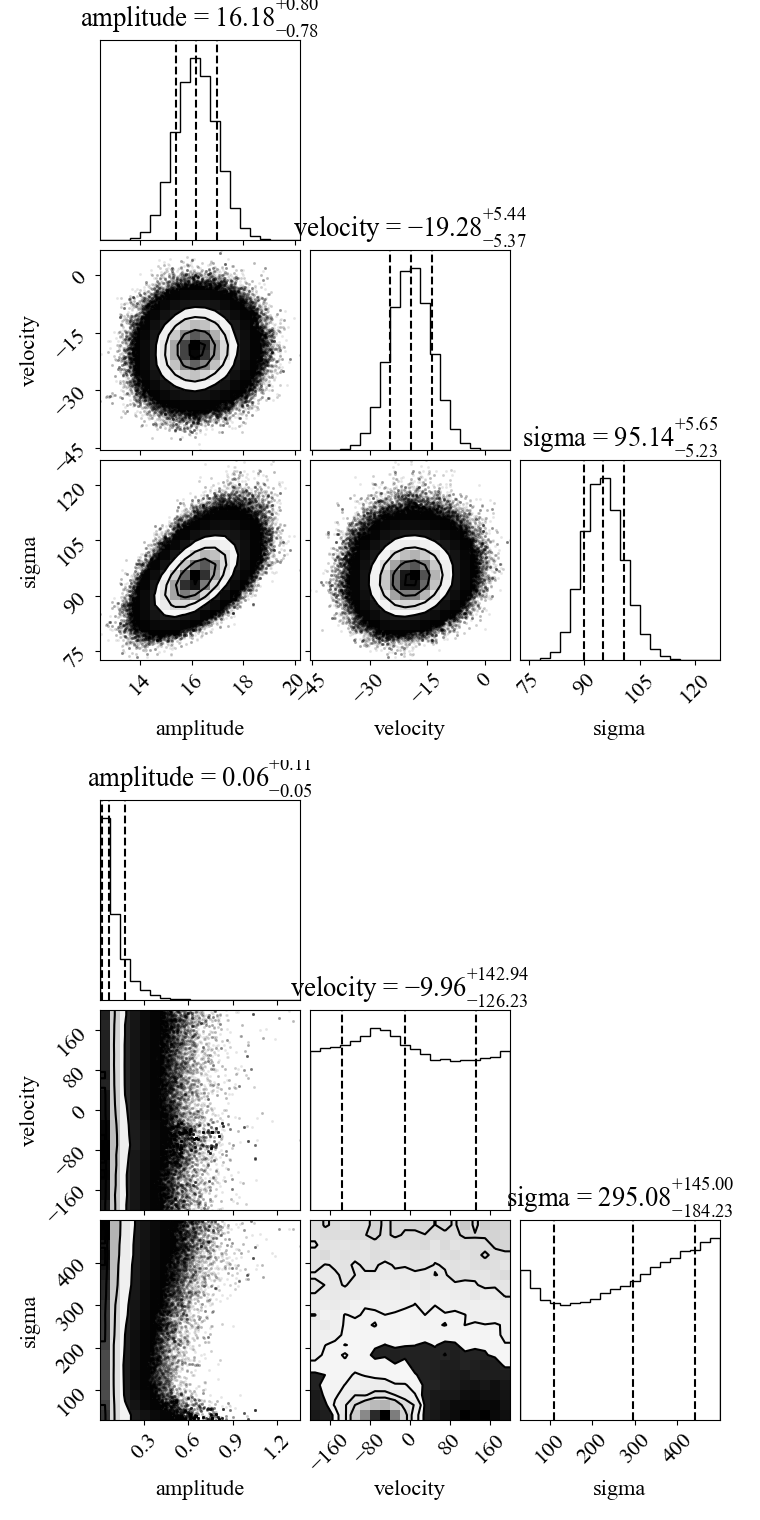}{0.2\textwidth}{(f)}
           \rotatefig{0}{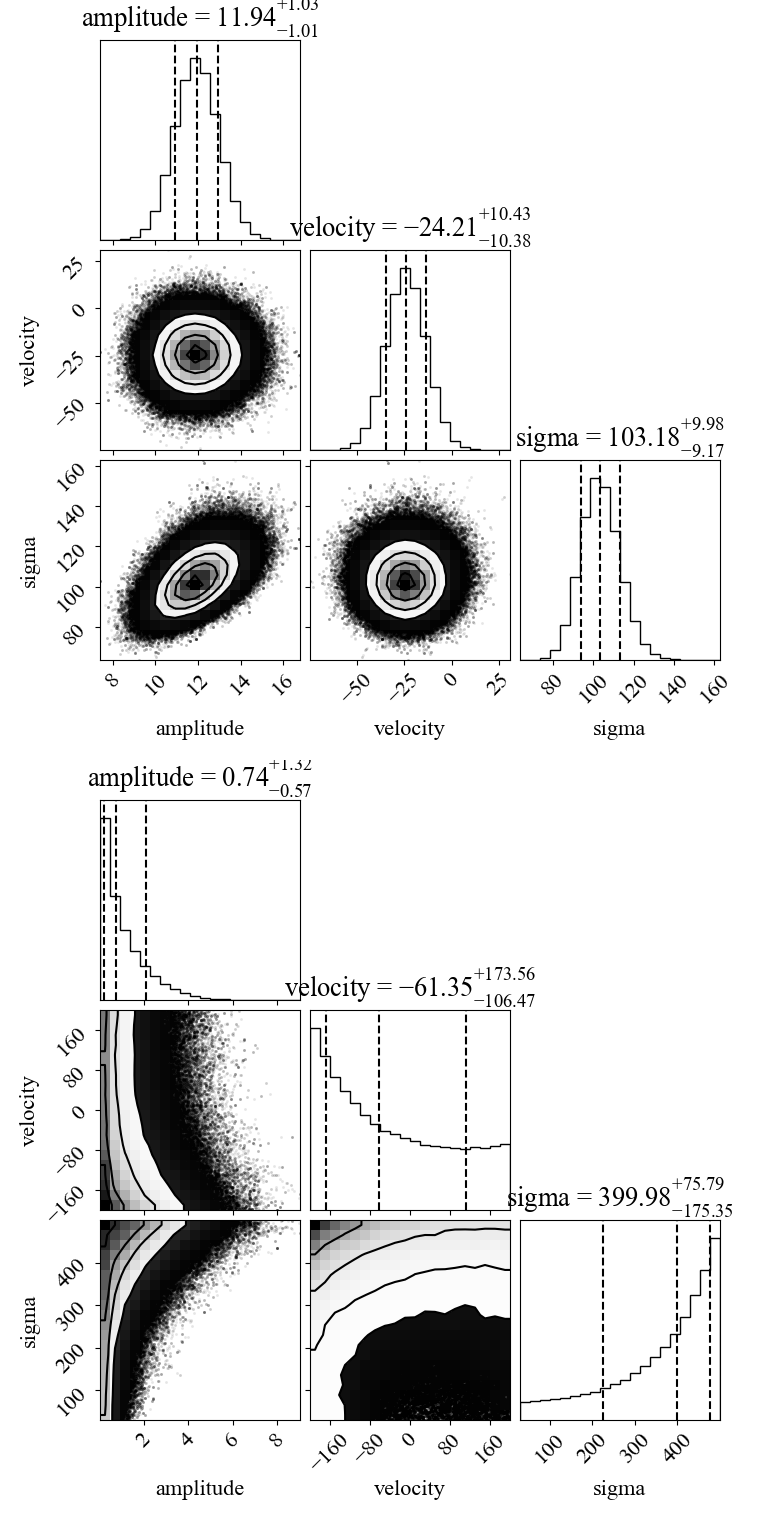}{0.2\textwidth}{(g)}
           \rotatefig{0}{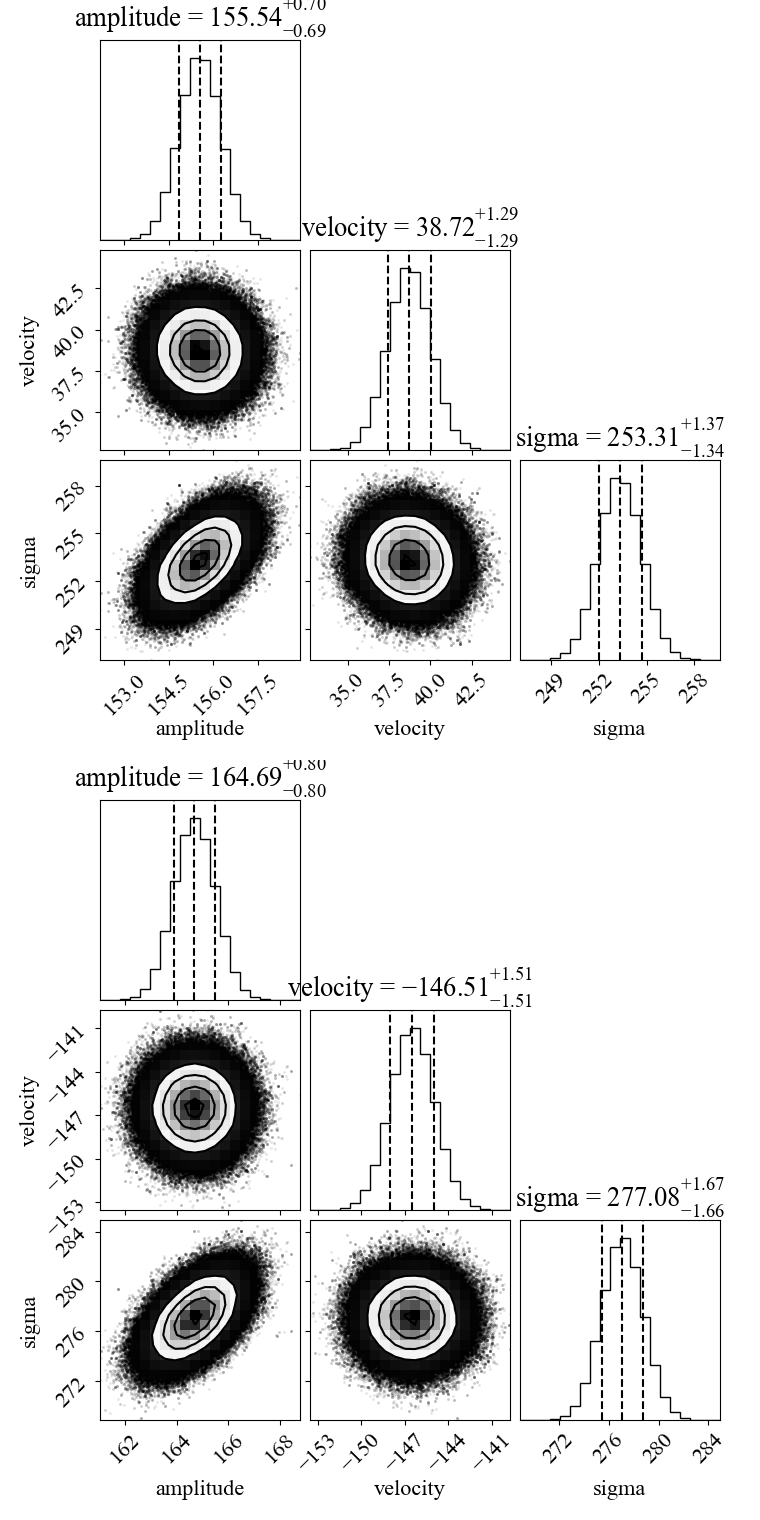}{0.2\textwidth}{(h)}
\rotatefig{0}{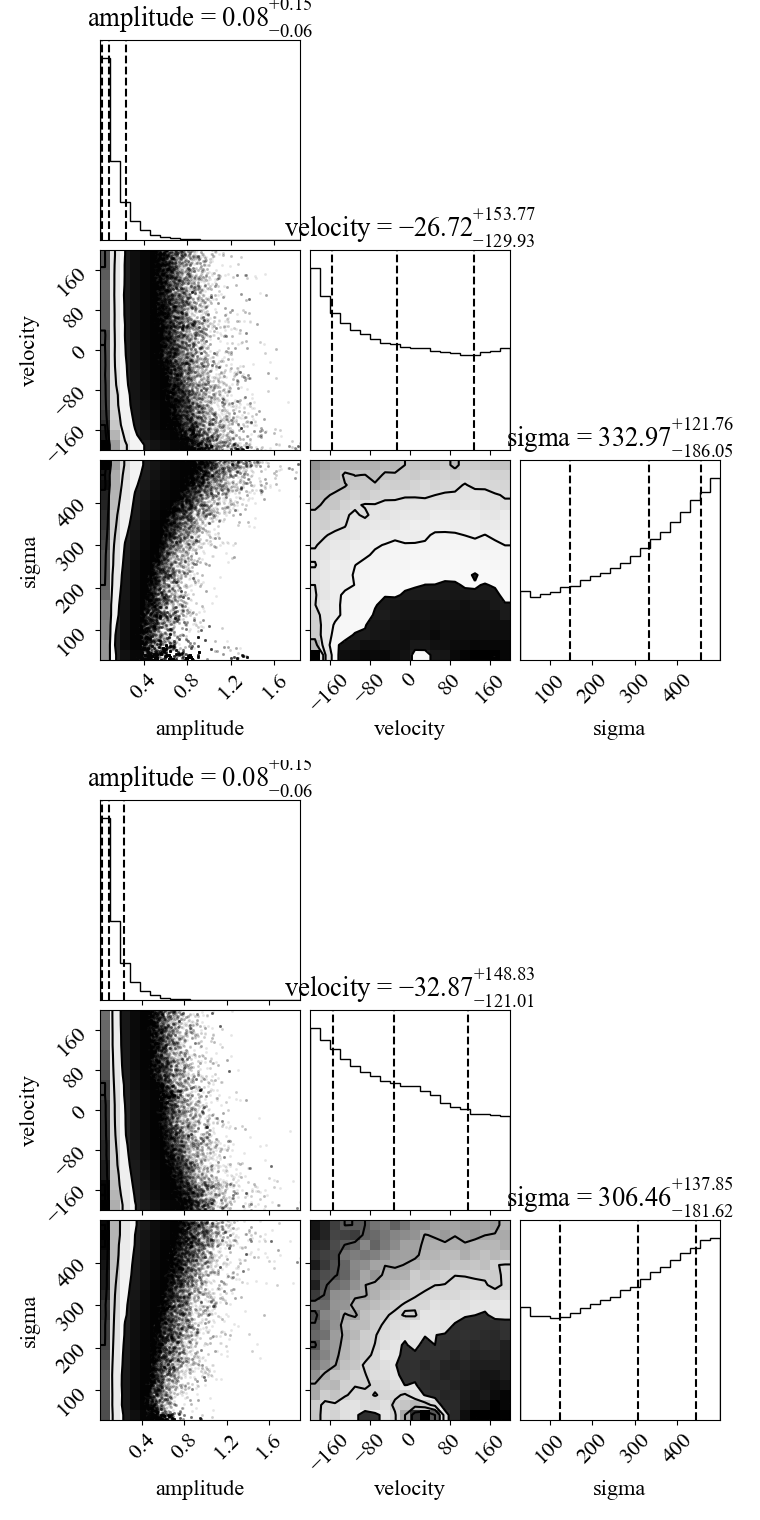}{0.2\textwidth}{(i)}
\rotatefig{0}{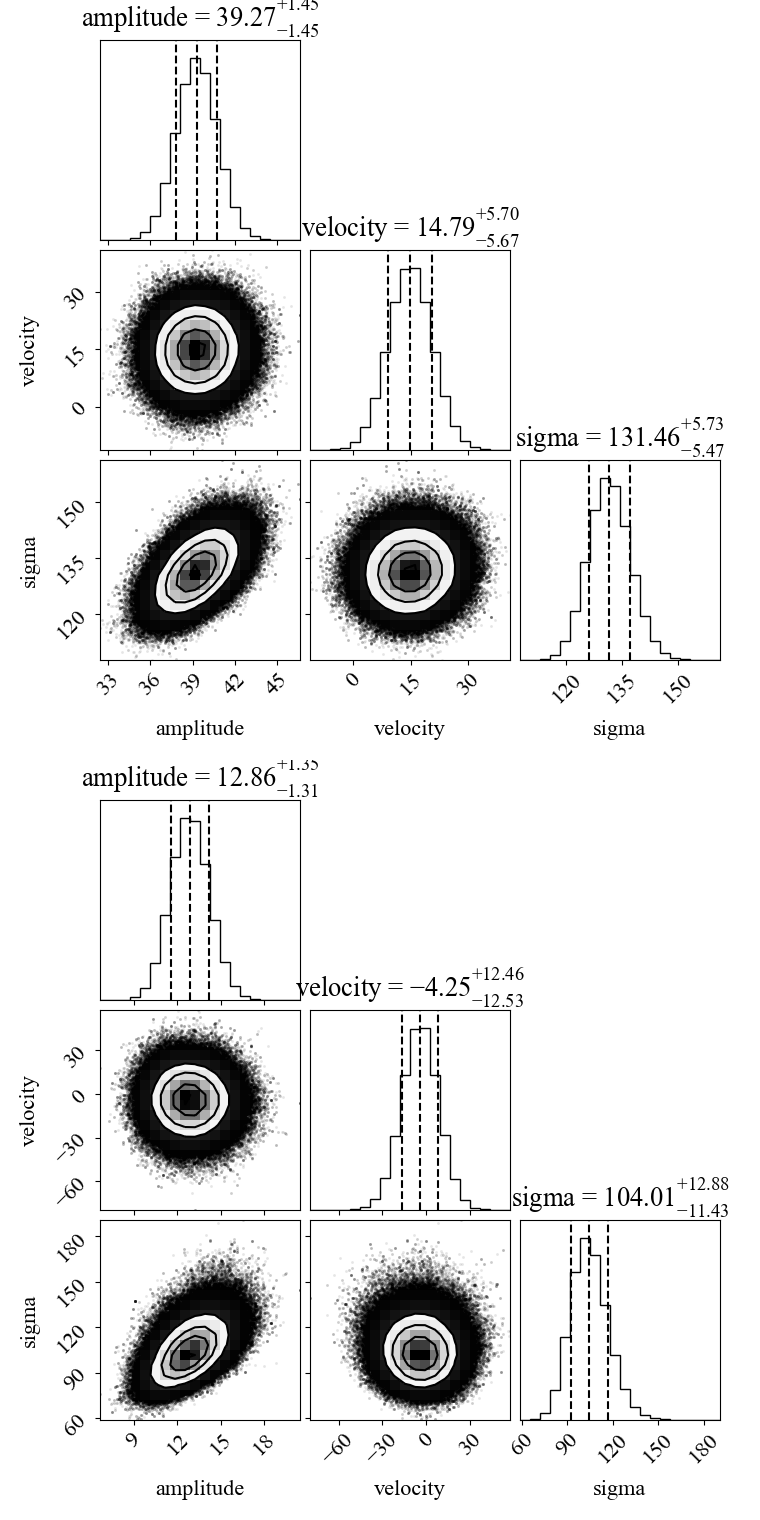}{0.2\textwidth}{(j)}
} 
\gridline{           
           \rotatefig{0}{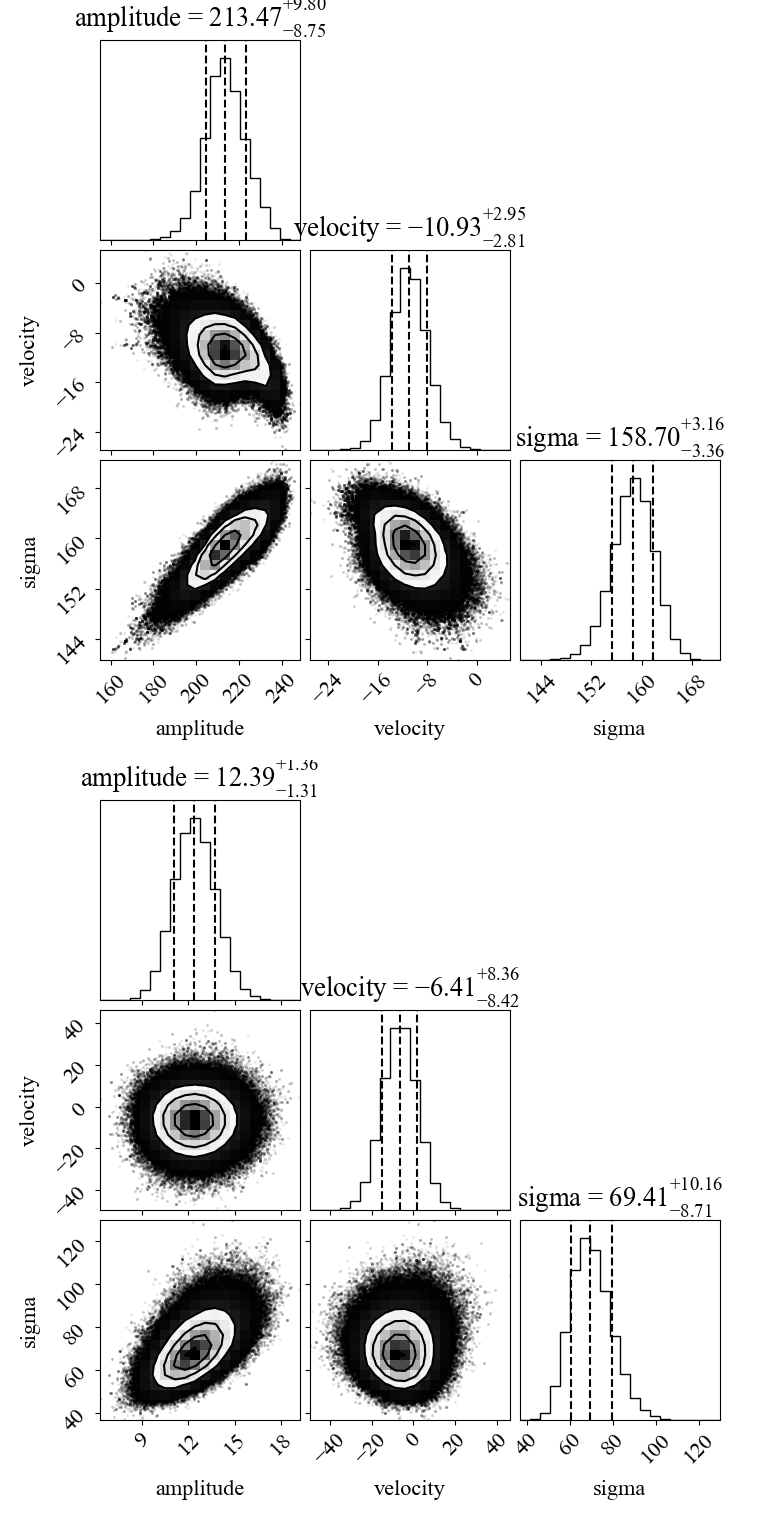}{0.2\textwidth}{(k)}
           \rotatefig{0}{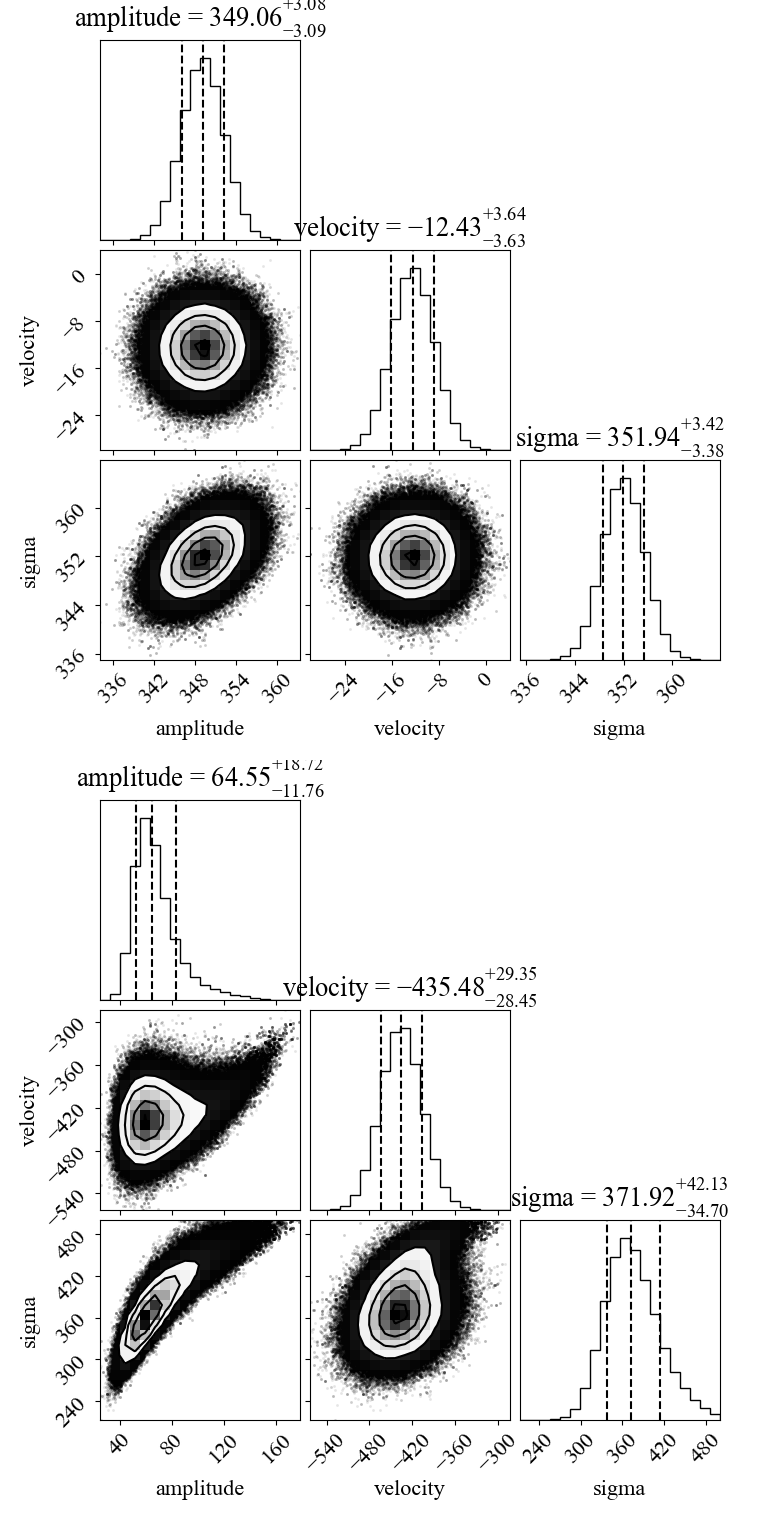}{0.2\textwidth}{(l)}
\rotatefig{0}{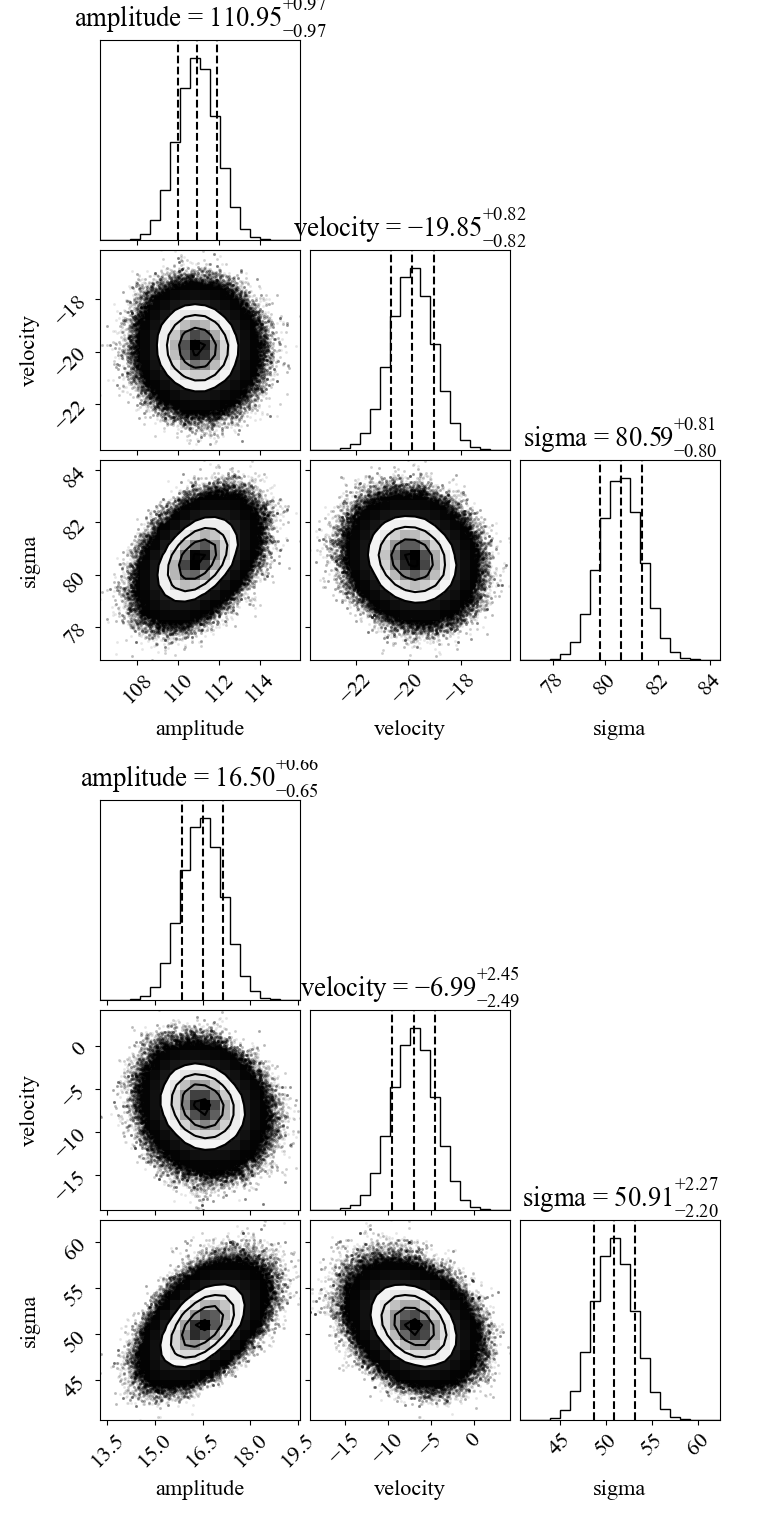}{0.2\textwidth}{(m)}
           }   
          \caption{Corner plots for the [\ion{Fe}{2}] (upper panel) and [\ion{P}{2}] (lower panel) line fitting in Figure \ref{fig:lineprofile}. See the caption of Figure \ref{fig:lineprofile} for the target name.}
\label{fig:contour}
\end{figure*}

We fit the [\ion{Fe}{2}] and [\ion{P}{2}] lines with Gaussians.
Figure \ref{fig:contour} displays the corner plots in the MCMC fitting shown in Figure \ref{fig:lineprofile}. The Gaussian amplitudes, central velocity, and velocity dispersion are shown. For [\ion{Fe}{2}] in NGC 4507 and [\ion{P}{2}] in NGC 6240, the results for the Gaussian with the largest EW are only shown.

\bibliography{01}{}
\bibliographystyle{aasjournal}



\end{document}